\documentclass[aps,twocolumn,floats,nofootinbib]{revtex4}
\usepackage{graphics,graphicx,epsfig}
\usepackage{amssymb,color}
\usepackage{epsf,epstopdf}
\usepackage {amsmath}
\usepackage[rightcaption]{sidecap}

\newcommand{\beq}{\begin{equation}}
\newcommand{\eeq}{\end{equation}}
\newcommand{\beqn}{\begin{eqnarray}}
\newcommand{\eeqn}{\end{eqnarray}}

\sidecaptionvpos{figure}{t}

\begin{document}

\title{Optimal decoding of information from a genetic network}

\author{Mariela D. Petkova,$^{a,b,e}$  Ga\v{s}per Tka\v{c}ik,$^f$  William Bialek,$^{a,b}$ Eric F. Wieschaus,$^{b,c,d}$  and Thomas Gregor$^{a,b}$}

\affiliation{$^a$Joseph Henry Laboratories of Physics, $^b$Lewis--Sigler Institute for Integrative Genomics, $^c$Department of Molecular Biology,  and $^d$Howard Hughes Medical Institute, Princeton University, Princeton NJ 08544\\
$^e$Program in Biophysics, Harvard University, Cambridge MA 02138\\
$^f$Institute of Science and Technology Austria, Am Campus 1, A--3400 Klosterneuburg, Austria}

\date{\today}

\begin{abstract}
Gene expression levels carry information about signals that have functional significance for the organism. Using the gap gene network in the fruit fly embryo as an example, we show how this information can be decoded, building a dictionary that translates expression levels into a map of implied positions.  The optimal decoder makes use of graded variations in absolute expression level, resulting in positional estimates that are precise to $\sim 1\%$ of the embryo's length.  We test this optimal decoder by analyzing gap gene expression in embryos lacking some of the primary maternal inputs to the network.  The resulting maps are distorted, and these distortions predict, with no free parameters, the positions of expression stripes for the pair-rule genes in the mutant embryos.   
\end{abstract}

\maketitle

\section{Introduction}

Biological networks   transform multiple input signals into outputs that have functional significance for the organism.  One path to understanding these transformations is to  ``read out,'' or decode this relevant information directly from the network activity. In neural networks, for example, features of the organism's sensory inputs and motor outputs have been decoded from observed action potential sequences, sometimes with very high accuracy~\cite{spikes,decode2,marre+al_14}.  Decoding provides an explicit test for hypotheses about how biologically meaningful information is represented in the network and which computations are needed to recover it. Here we address these questions in a small genetic network, taking advantage of experimental methods that allow us to measure, quantitatively, the simultaneous expression levels of multiple genes.

\begin{figure*}
\centerline{\includegraphics[width = \linewidth]{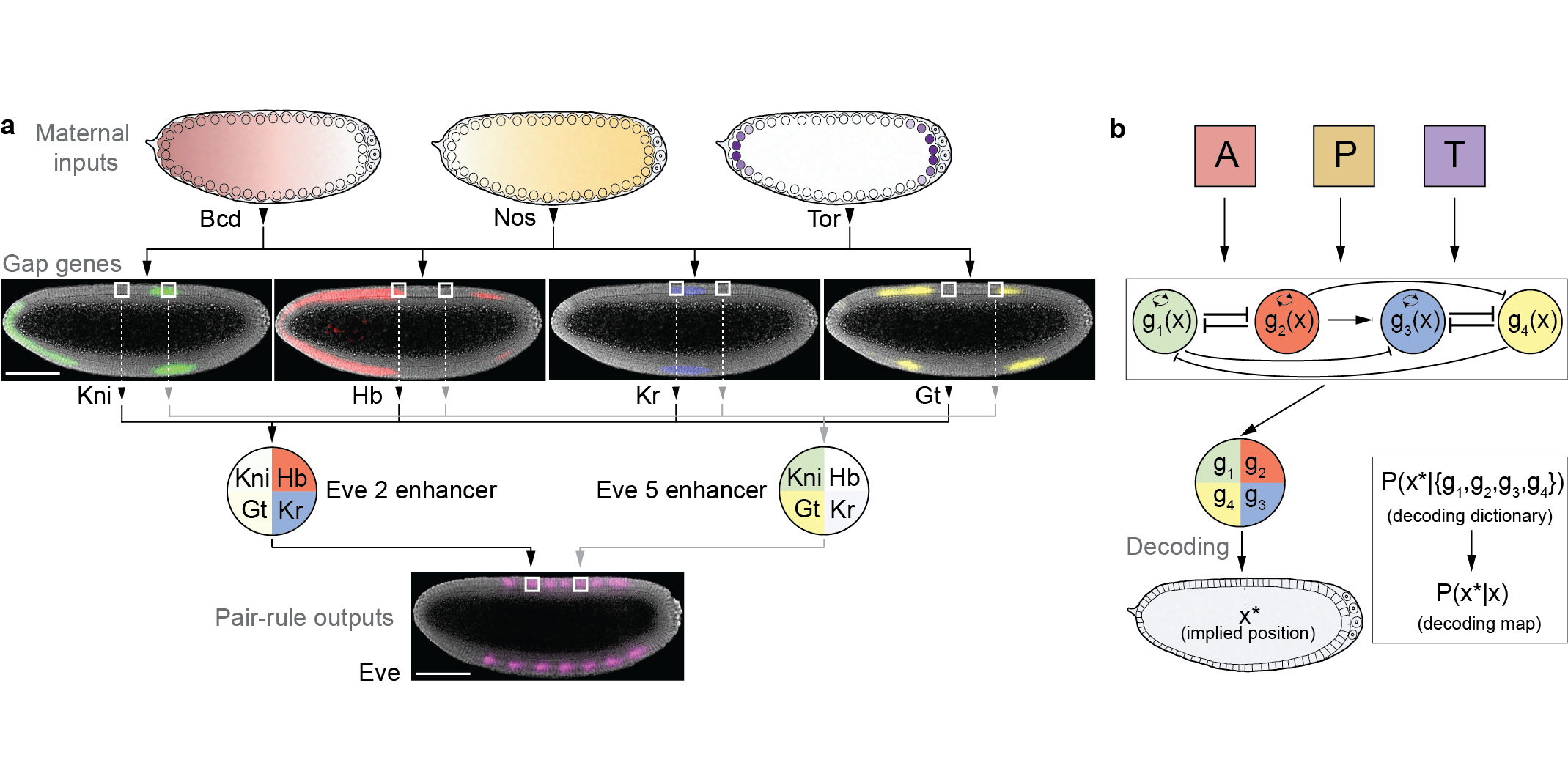}}
\caption{{\bf Decoding in a genetic patterning network.} {\bf a.} In the early {\it Drosophila} embryo, maternally provided morphogens ({\it bcd, nos, tor}) regulate the expression of gap genes ({\it kni, kr, gt, hb}), shown here in a  mid--sagittal slice through an embryo during nuclear cycle 14.  Enhancers (schematically depicted as circles) respond to combinations of gap gene concentrations to drive pair-rule gene expression, which occurs in a  striped pattern.  Each stripe appears with a precise and reproducible  position. {\bf b.} Schematic depiction of the anterior-posterior patterning network in the early embryo. The network can be viewed as an input/output device that encodes physical location $x$ in the embryo using concentrations of gap gene proteins, $g_1, g_2, g_3, g_4$.  Optimal decoding is a well-posed mathematical problem, solved by selecting the best estimate of position $(x^*)$ from the ``decoding dictionary,'' $P(x^*|\{g_i\})$.
\label{F1}}
\end{figure*}

The gap genes involved in patterning the early fruit fly embryo provide a particularly accessible example of the decoding problem  \cite{dev_general,gergen+al_86,lawrence_92,jaeger_11,briscoe+small_15}.   As schematized in Fig \ref{F1}a, the gap genes form an interconnected layer in an otherwise feed-forward network that takes inputs from the primary maternal morphogens to drive the expression of the pair-rule genes.     Pair-rule expression occurs in stripes that are precisely and reproducibly positioned within the embryo, forming an outline for the segmented body plan of the fully developed organism.  In this system,  states of the network are the expression levels of the gap genes, and the functional  information being encoded is  the position of cells along the anterior--posterior axis (Fig \ref{F1}b).

We will use the gap gene network  to distinguish between two very different views of network function.  In one view, the control of gene expression is noisy, network interactions are essential to achieve robustness against uncontrolled variations in the input signals, and many layers   are needed to funnel the states of cells toward reliable choices among discrete fates (canalization).  In this view, information available to individual cells from gap gene expression levels should be very limited, and the accuracy associated with a reproducible body plan should emerge only in subsequent layers of processing.  In another view, noise levels are as low as possible given the limited number of molecules involved \cite{gregor+al_07b}, the reproducibility of developmental patterning can be traced back to reproducible maternal inputs \cite{petkova+al_14}, and network interactions are selected to extract the maximum amount of information from these inputs \cite{tkacik+al_08a,genesI,genesII,genesIII,sokolowski+tkacik_15,sokolowski+al_16}.  In support of this latter, precisionist view, we have shown that the expression levels of the four gap genes, taken together at one moment in time, provide enough information to specify the location of each  cell  along the anterior--posterior axis of the embryo with $\sim 1\%$ precision \cite{dubuis+al_13b,tkacik+al_15}, comparable to the precision with which pair-rule  stripes and other morphological markers are specified.  

Showing that precise information is available  is not enough: to say that the system operates in a precisionist mode we must show that this information is  used by the embryo to guide subsequent developmental events.   To do this, we will decode the information carried by gap gene expression levels, and manipulate these signals through mutations in the maternal inputs to the network.  If the organism really uses the information that we have found encoded in the gap gene network, then our decoding of this information should make quantitative predictions for the functional consequences of mutations at the input.

In the fly embryo, information encoded in the levels of gap gene expression is decoded by the complex regulatory logic of the pair rule gene enhancers  \cite{small+al_91,levine_10}, and one might think that decoding requires us to make a model of these molecular events.  But if the embryo makes optimal use of the available information \cite{gregor+al_07b,tkacik+al_08,dubuis+al_13b}, then the algorithm for decoding positional information   is determined, mathematically, by the pattern of gene expression as a function of position and by the distribution of fluctuations around this mean.  Exploiting our ability to make precise, quantitative measurements on the expression of all four gap genes simultaneously  \cite{dubuis+al_13a}, we can characterize these fluctuations and construct the optimal decoder with no free parameters; this is not a model of the decoder, but rather a theoretical prediction of what the decoder should be if the embryo performs optimally.     Applied to gap gene expression levels in wild-type embryos, this optimal decoder generates unambiguous estimates of position, which fluctuate only slightly from  the true position.  In mutant embryos deficient for various primary maternal inputs, gap gene expression changes at each point in the embryo, and the decoder generates distorted ``maps'' of implied vs true position;  these maps in turn can be used to predict the locations of pair-rule stripes.  We find that these predictions are in detailed, quantitative agreement with experiment, again with no adjustable parameters. These results strongly support the idea that biologically relevant information is carried by precisely controlled gene expression levels, so that the function of this genetic network is fundamentally quantitative, down to $\sim 1\%$ accuracy in the encoded variable.

\section{Dictionaries and maps}

\begin{figure*}
\centerline{\includegraphics[width = \linewidth]{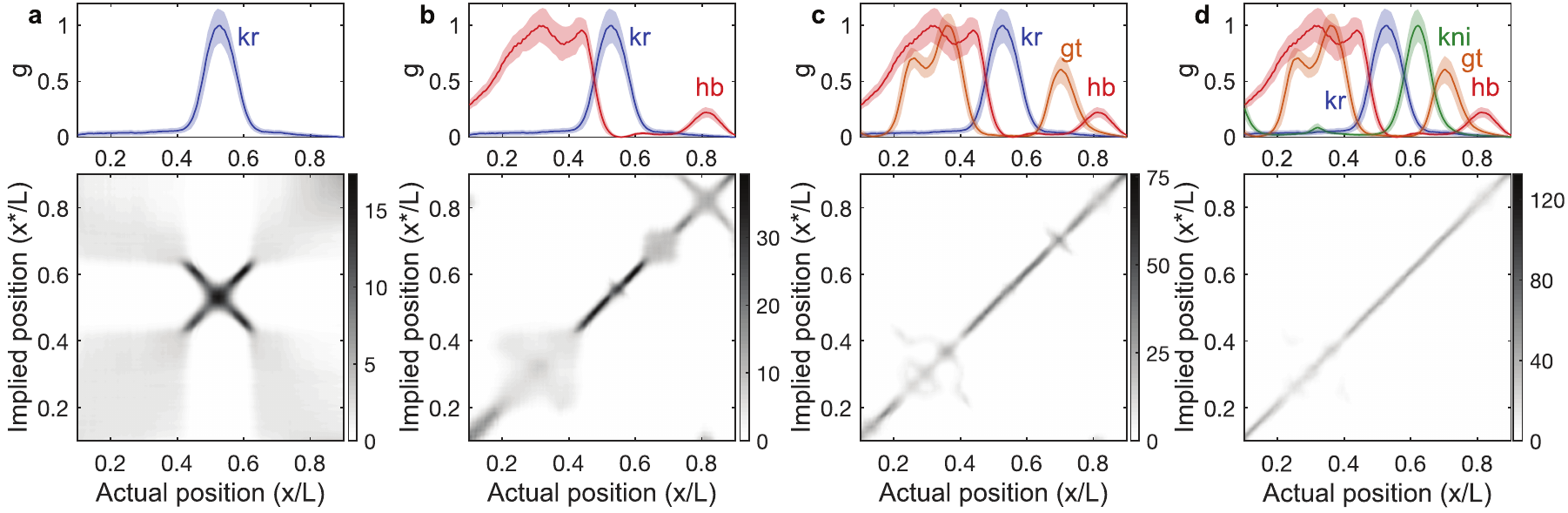}}
\caption{{\bf Refinement of decoding precision with increasing number of gap genes in WT embryos.} Top row: dorsal fluorescence intensity profile(s) from 38 embryos 38--48 min into nuclear cycle 14 (mean $\pm$ SD); units scaled so that 0 (1) corresponds to minimum (maximum) mean expression. Bottom row: average decoding maps representing the probability density $P(x^*|x)$; each decoding map is constructed by averaging the individual probability density distributions of the 38 embryos (see also Fig.~\ref{SF0}). {\bf a.}  Decoding using single gene (Kr, blue) can be imprecise and ambiguous (0.1$\!<\!x/L\!<\!$ 0.43 and 0.63 $\!<\!x/L\!<\!$ 0.9), precise but ambiguous (0.43 $\!<\!x/L\!<\!$ 0.5 and 0.55 $\!<\!x/L\!<\!$ 0.63), and imprecise but unambiguous (0.5 $\!<\!x/L\!<\!$ 0.55), see also Fig.~\ref{SF2}.
{\bf b.} Decoding using a combination of two genes, Kr (blue) and Hb (red), see also Fig.~\ref{SF3}.  {\bf c.} Decoding using three genes, Kr (blue), Hb (red), and Gt (orange), see also Fig.~\ref{SF4}.  {\bf d.} Decoding using all four gap genes, which is precise and almost perfectly unambiguous along the whole length of the embryo.  
\label{F2}} 
\end{figure*}

The expression levels of the gap genes vary with position, but even at a single position there are variations around the mean expression level.  Thus, at each point $x$ along the anterior--posterior axis of the embryo,  there is a probability distribution,  $P(\{g_{\rm i}\}|x)$, where $\{g_{\rm i}\} = g_1, \, g_2,\, \cdots $ are the expression levels of the genes, in our case the gap genes {\em hunchback,} {\em kr\"uppel,} {\em knirps,} and {\em giant}.   The problem facing a cell, however, is not to predict the expression level given knowledge of its position.  On the contrary, the cell must respond to the expression level by taking actions appropriate to its position.  Mathematically, the possible positions of the cell are  drawn out of the conditional distribution
\begin{equation}
P(x^*|\{g_{\rm i}\}) = \frac{P(\{g_{\rm i}\}|x^*) P_X(x^*)}{P_G (\{g_{\rm i}\})},
\label{decode1gene}
\end{equation}
which we construct using Bayes' rule; we use $x^*$ to remind us that this is the position {\em implied} by the expression levels, rather than the actual position of the cell. $P_X(x^*)$ is the probability that a cell will be at position $x^*$, independent of gene expression level, and in the simplest case the cells are uniformly distributed along the axis $0 < x \leq L$, so that 
$P_X(x^*) = 1/L$.   $P_G(\{g_{\rm i}\})$ is the probability that we will find a cell with expression levels $\{g_{\rm i}\}$, independent of its position,     $P_G (\{g_{\rm i}\}) = \int dx\, P(\{g_{\rm i}\}|x) P_X(x)$.

The probability distribution $P(x^*|\{g_{\rm i}\})$ contains everything that a cell could ``know'' about its position based on the expression levels $\{g_{\rm i}\}$.  If the cell responds to the expression levels in a way that matches the structure of this distribution, then that response is part of an optimal decoding.   To be more precise, if the probability distribution $P(x^*|\{g_{\rm i}\})$ has a single, reasonably sharp peak at $x^* = X^*(\{g_{\rm i}\})$, then  we can translate expression levels back into positions, unambiguously, using a dictionary $\{g_{\rm i}\} \rightarrow X^*$, and the width of the distribution $P(x^*|\{g_{\rm i}\})$ tells us how much error or uncertainty there will be in this estimate of position.  We could formalize this by asking for the value of $x^*$ that maximizes $P(x^*|\{g_{\rm i}\})$, which is called the maximum a posteriori estimator, or for the value of $x^*$ such that we make smallest mean-square error relative to the true position, and these are essentially the same if there is a single sharp peak in the distribution.  But it also is possible that   $P(x^*|\{g_{\rm i}\})$ has multiple peaks, in which case there are genuine ambiguities in decoding, and hence the gene expression levels we are analyzing in fact are not sufficient to determine the position and hence the fate of the cell.  Since we don't know in advance that decoding will be successful, it is useful to keep track of the entire distribution.  Indeed, since   the information that we are discussing is information about position, it is useful to express the results of decoding as a {\em map}.  

Consider a particular embryo $\alpha$, in which we find expression levels $\{g_{\rm i}^\alpha (x )\}$ in the cell at actual position $x $.    Instead of thinking of the distribution of possible positions in Eq (\ref{decode1gene}) as depending on all of the gene expression levels, we can think of it as depending on the actual position of the cell, forming a map of possible or inferred positions vs actual positions,
\begin{equation}
P_{\rm map}^\alpha (x^* | x ) = P(x |\{g_{\rm i}\}){\bigg |}_{\{g_{\rm i}\} = \{g_{\rm i}^\alpha (x )\}} .
\label{map1}
\end{equation}
If the genes we are considering provide enough information to specify position accurately and unambiguously, then $P_{\rm map}^\alpha (x^*  | x )$ will be a narrow ridge of density surrounding the line such that the implied position is equal to the true position, $x^* = x$.

To implement these ideas, we begin by measuring expression of all four gap genes, simultaneously,   as described in detail in Appendix \ref{app_exp}.  We sort the emrbyos by time, and focus on a window   38$\!\--\!$48 min into nuclear cycle fourteen, during which gap gene expression provides the most information about position \cite{dubuis+al_13b,dubuis+al_13a,tkacik+al_15}.   We approximate the distribution $P(\{g_{\rm i}\}|x)$ as being Gaussian \cite{dubuis+al_13b}, which means that we are estimating the mean expression level at each position, across the ensemble of embryos, and the covariance matrix (Fig \ref{SF12}) that describes fluctuations around this mean. 
Given these features of the data, the construction of the decoding map proceeds with no adjustable parameters; for   details see Appendix \ref{methods+thy} and Figs \ref{SF0} through \ref{SF4}.

In Figure \ref{F2}  we show the decoding maps based on the information carried by one, two, three, or all four gap genes.   For most locations in the embryo, decoding based on a single gene provides   little information, as for Kr in Fig \ref{F2}a and \ref{SF0}; other examples are in Fig \ref{SF2}.  In small regions of the embryo, decoding can be  more precise, but substantial ambiguities remain, where one   expression level is equally consistent with two different implied positions.  Decoding based on two (Figs \ref{F2}b and \ref{SF3}) or three (Figs  \ref{F2}c and  \ref{SF4}) genes results in   less  ambiguity and more precision; we can  follow this sharpening of the distribution  by the increasing height of its peak, since narrower distributions   have higher peaks to maintain normalization, or by estimating the median width of the distribution $P_{\rm map}^\alpha(x^* | x  )$ (Fig \ref{SF11}d).  Finally,  with all four genes,   the distribution $P_{\rm map}^\alpha(x^* | x  )$ is approximately Gaussian, with a width $\sigma_x \approx 0.01L$, for nearly all points $x$.     

It has been conventional to think about a single morphogen gradient coding for position \cite{wolpert_69}, or about ``expression domains'' of single gap genes pointing to large regions of the embryo. Figure \ref{F2} shows how multiple expression levels combine, quantitatively, to synthesize an unambiguous code for position that reaches extraordinary precision.  This one percent precision in the encoding of position is also the precision with which  subsequent developmental markers, including the pair-rule gene stripes and the cephalic furrow, are generated \cite{dubuis+al_13b,tkacik+al_15}.

\section{Testing the dictionary}

\begin{figure*}
\centerline{\includegraphics[width = \linewidth]{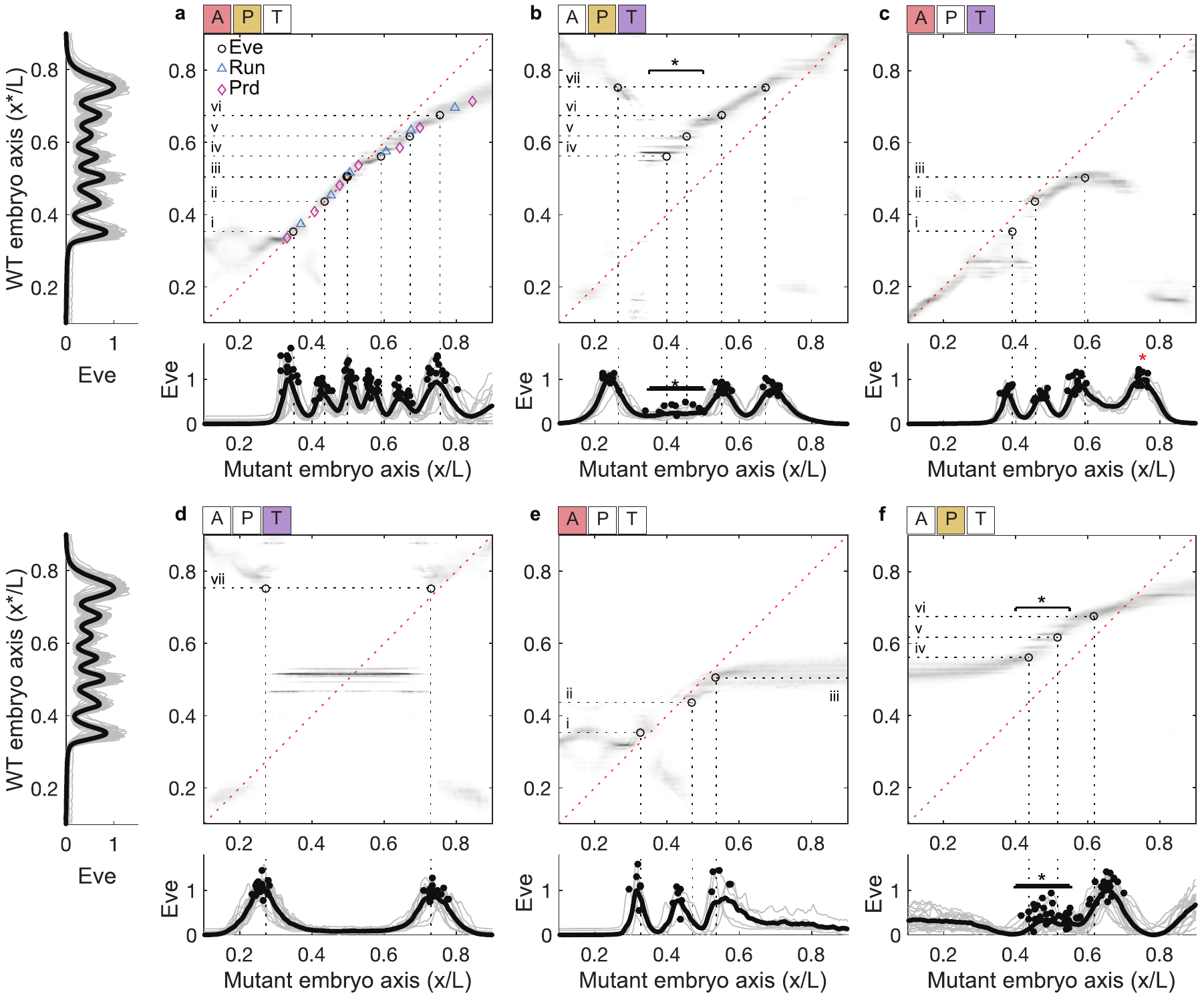}}
\caption{{\bf Using decoding maps to   predict stripe locations in mutant embryos.} Average decoding maps for six maternal mutant backgrounds: {\bf a.} {\it etsl};  {\bf b.} {\it bcd$^{\rm E1}$}; {\bf c.}  {\it osk};  {\bf d.} {\it bcd$^{\rm E1} $osk}; {\bf e.} {\it osk tsl}; {\bf f.} {\it bcd$^{\rm E1} $tsl}. The average locations of wild-type Eve stripes (horizontal dotted lines) are used to predict Eve stripes in the mutant backgrounds: we expect to observe stripes at locations along the mutant embryo axis, where the horizontal dotted lines intersect the peak of the probability density (open black circles and vertical dotted lines). Measured Eve expression profiles in wild-type embryos (left side of {\bf a} and {\bf d}), and in mutant embryos (below the corresponding decoding map);  individual profiles (gray), mean profile (black), and peak locations (black dots).    Horizontal starred bars (panels {\bf b} and {\bf f}) indicate locations where the expressed number of Eve stripes is variable (see Fig.~\ref{SF14} and Fig.~\ref{SF13} for examples). When the horizontal lines intersect a broad probability distribution, we expect to observe diffuse Eve stripes as in {\bf e} (see Fig.~\ref{SF14} for example). The red star in {\bf c} shows an observed Eve stripe which is not predicted by the decoding map. Panel {\bf a} shows additional predictions for Run (cyan) and Prd (magenta) stripes, which also agree with measured peak positions (see also Figs \ref{SF6} and \ref{SF7}).
\label{F3}} 
\end{figure*}

The fact that the four gap genes carry precise, unambiguous information about position does not mean that the embryo uses this information.  To test whether this is the case, we perturbed the maternal signals Bicoid ({\em bcd}), Nanos ({\em nos}), and Torso-like ({\em tsl}), which strongly affect the output of the gap gene network (Fig \ref{SF1}).   If our optimal decoding strategy is used by the  embryo, our decoder should generate meaningful position estimates in the mutants. In particular, we will use the stripes of pair-rule gene expression as a measure of the embryo's readout of positional information, and compare this with our own.

We have analyzed embryos from lines in which we delete {\em bcd}, {\em tsl}, and {\em osk} (which controls the localization of the {\em nos} signal), singly and in pairs.  For each of these six possibilities, we have measured expression levels for all four gap genes simultaneously, in samples that also include wild-type embryos; results are summarized in Fig \ref{SF1}.  In every case, we construct the dictionary for decoding gap gene expression levels from the wild-type embryos, and then apply this dictionary to the mutants measured in the same batch, avoiding any concerns about systematic variations in staining, imaging, etc. across batches.  The results of these  analyses are a series of decoding maps, shown in Fig \ref{F3}, which  should be compared to the map for wild-type embryos in Fig \ref{F2}d.   

Many features of the decoding maps in Fig \ref{F3} are expected from previous, qualitative characterizations of these mutants.  Thus, when we delete {\em tsl}  the distortions are largely at the two ends of the embryo (Fig \ref{F3}a),   since expression of {\em tsl} is confined to the poles; when we delete {\em bcd} there are major distortions in the anterior portion of the map  (Fig \ref{F3}b), where the concentration of Bcd protein is highest; and  when we delete {\em osk} (Fig \ref{F3}c), we see major distortions in the posterior, consistent with {\em nos} being a posterior determinant.  More striking are the quantitative predictions that we obtain by decoding.

When we delete {\em bcd}, quantitative distortions of the map extend into the posterior half of the embryo, so that, for example, the expression levels found at $x /L =0.7$ generate a distribution $P_{\rm map}^\alpha(x^* | x ) $  in which the most likely value is $x^*/L = 0.75$, and at $x/L =0.55$ the most likely estimate is $x^*/L = 0.67$.  Thus, even in the posterior half of the embryo, the map is shifted, and the plot of $x^*$ vs $x$ (following the ridge of high probability in the map) does not have unit slope.  But in the wild-type embryo, the positions $x/L = 0.75$ and $x/L = 0.67$ are associated with peaks in the stripes of expression for the pair-rule gene {\em eve}, as shown at left in Fig \ref{F3}.  If the machinery for interpreting gap gene expression is using the same dictionary that we have constructed here, then we predict that the {\em bcd} deletion mutants should shift these {\em eve} stripes to $x/L = 0.7$ and $x/L = 0.55$, and this is what we see (Fig \ref{F3}b).    Even more dramatically, expression levels at $x/L =0.23$ in the {\em bcd} mutant are decoded as $x^*/L = 0.75$ with high probability, and correspondingly there is an {\em eve} expression stripe at this anomalously anterior location.   This is predicted to be not a displacement of the first (nearest) {\em eve}  stripe, but rather a duplication of the seventh stripe, which is consistent with classical observations on cuticle morphology in these mutants \cite{driever+al_88b}, and with recent RNAi/reporter experiments \cite{staller+al_15}.
 
The quantitative agreement between the decoding maps and the locations of the {\em eve} stripes  extends to all six examples of single and double maternal mutants shown in Fig \ref{F3}.  Notably, there is good agreement both when the shifts are small, as with the deletion of {\em tsl} (Fig \ref{F3}a), and when the shifts are much larger, resulting in the deletion of several stripes, as with the {\em bcd osk} and {\em osk tsl} double mutants (Figs \ref{F3}d and e).   In cases where the implied position of a stripe crosses a diffuse band of probability density in the map, as with {\em eve} stripe iii in the {\em osk tsl} mutant (Fig \ref{F3}e), we might expect that there would be expression of {\em eve} but not a sharp stripe, and this is what we see; similar effects occur in the anterior of the {\em bcd tsl} mutant (Fig \ref{F3}f). 

For simplicity Fig \ref{F3} shows decoding maps that are averaged over all embryos for each mutant line.  If we focus on decoding maps for individual embryos instead, their variability predicts the embryo-to-embryo variability in pair-rule gene expression. In particular, for {\em bcd tsl} mutants the positions that map to the wild-type locations of {\em eve} stripes four and five ($x^*/L = 0.56$ and $x^*/L = 0.62$) vary substantially in the window $0.4 < x/L  < 0.6$.  If we look at the {\em eve} expression patterns in individual embryos (thin lines at bottom of Fig \ref{F3}f; for detailed analysis see Fig \ref{SF14}), we see two peaks  with variable positions, as predicted.   For the {\em bcd} mutant, the average decoding map again has density at $x^*/L = 0.56$ and $x^*/L = 0.62$ (Fig \ref{F3}b and Fig \ref{SF13}), but when we decode the gap gene expression patterns from individual mutant embryos we find that these features vary not only in their position but even in their presence or absence, so that individual embryos are predicted to have a variable number of {\em eve} stripes, and this is what we see.

We have repeated the analysis of Fig \ref{F3}, comparing our decoding maps with the expression profiles of the pair-rule genes {\em paired} ({\em prd}) and {\em runt} ({\em run}), as summarized in Figs \ref{SF6} and \ref{SF7}.  In most cases we  see relatively few stripes, which agrees with the theory since there are few intersections between expected stripe positions and the ridge of maximal probability.  In the case of the {\em tsl} mutant, however, the three different pair-rule stripes provide a rather dense collection of markers through the range $0.3 < x/L < 0.85$, as shown in detail in Fig \ref{F3}a.  These markers  trace the predicted ridge of maximal probability with very high accuracy.

\begin{figure}
\centerline{\includegraphics[width = \linewidth]{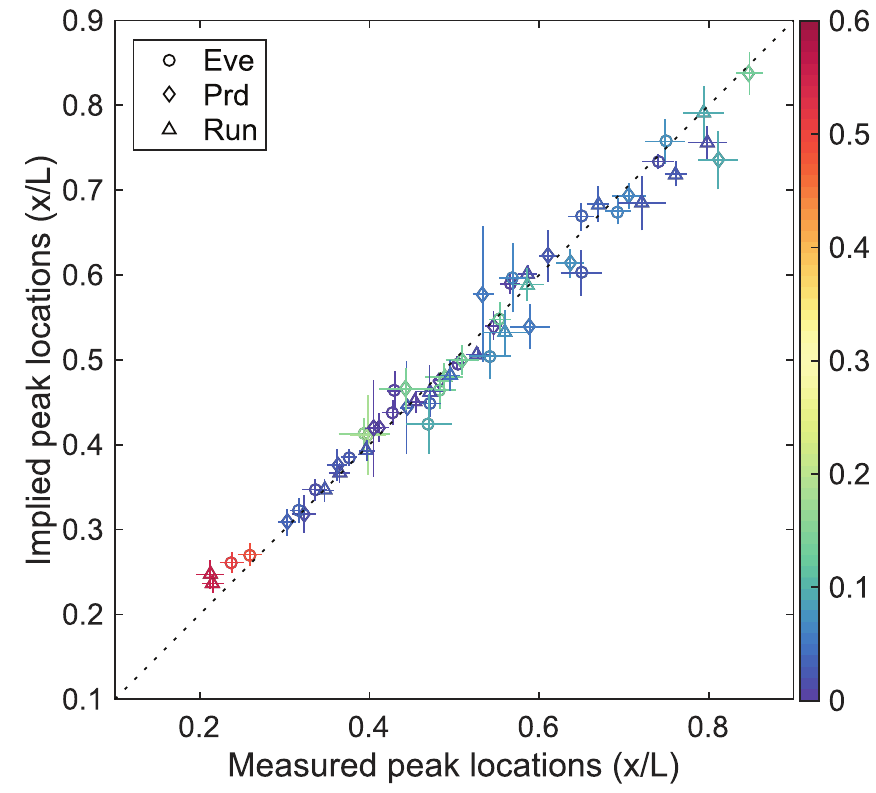}}
\caption{{\bf  Implied vs observed locations of pair-rule stripes in mutant embryos.}  Color scale (right) indicates the displacement of the observed peak in the mutant from its wild-type location (units, $x/L$). Horizontal axis: pair-rule stripe positions in mutant embryos, mean $\pm$ s.d. across embryos of a given genotype. Vertical axis: predictions from decoding the gap gene expression levels in mutant embryos, mean $\pm$ s.d. across embryos of a given genotype. We predict and observe a total of 57 stripes. Additionally, we predict and observe 9 diffuse stripes, which cannot be quantified by a single peak location (Appendix \ref{app:stripes}).  We observe, but do not predict 3 stripes; and predict, but do not observe 2 stripes. \label{summary}}
\label{F4}
\end{figure}

The results for all the {\em eve}, {\em run}, and {\em prd} stripes in all six mutants are summarized in Fig \ref{summary}.  For almost all the 57 stripe positions shown, the predicted position agrees with the measured position within the error bar defined by the standard deviation of these positions across embryos in our sample.  More subtly, when we decode the positional signals from multiple embryos, we find variations in the resulting maps, as noted above, and the standard deviation of predicted stripe positions are in most cases close to the observed variations (vertical and horizontal error bars, respectively).

We do note a small number of errors in our predictions.  In the {\em osk} mutants we observe a posterior Eve stripe where none is predicted (Fig \ref{F3}c), and in {\em bcd osk} mutants, a variable number of Prd stripes (Fig \ref{SF6}d). We predict a Run stripe at $x/L\sim 0.6$ where none is observed (Fig \ref{SF7}c); we have no prediction for the very blurred band of Run expression  at $x/L>0.7$ (Fig \ref{SF7}c). In addition, in the {\em bcd, tsl} mutant we predict a Run stripe at $x/L\sim 0.45$ where none is observed (Fig \ref{SF7}f).  This last failure occurs at a rare point where the combinations of gap gene expression are outside the range that we have sampled in the wild-type embryos (Fig \ref{SF5}), and thus we may be simply extrapolating the probability distributions too far.  The errors in the {\em osk} mutants cluster around a point where the predicted maps have a discontinuity, and thus are especially sensitive to our assumption that decoding is done in a purely local fashion.  A small amount of spatial averaging, perhaps to reduce noise introduced in the expression of the pair-rule genes, would result in very different predictions.

\section{Discussion}

The expression levels of genes carry information about signals that matter in the life of the organism.  We have tried to answer several questions:  How much information is available \cite{dubuis+al_13b}?  What features of the expression pattern are crucial in conveying this information?  How can the organism ``read out'' what is relevant?

In the developing embryo, one biologically relevant signal is position.  In trying to understand how positional information is represented, we have focused on gene  expression levels at a single moment in time, and on the levels observable by a single cell.  It is possible that the cells could extract  more reliable information by averaging expression levels over time, although  since protein concentrations accumulate the snapshot that we consider already reflects some degree of temporal averaging, and further averaging may be of limited impact.  Similarly,  individual cells might be able to extract more information by communicating with their neighbors \cite{gregor+al_07b,erdmann+al_09,sokolowski+tkacik_15}, but we know that fluctuations in gap gene expression are correlated over substantial distances \cite{krotov+al_14}, limiting the benefits of such spatial averaging.    Importantly, once we focus on the information available locally in space and time, our experimental tools make it possible to give an essentially complete characterization of the available information and  the nature of its encoding, as summarized by the probability distributions $P(\{g_{\rm i}\}|x)$.

Positional information encoded by the gap gene expression levels is effectively decoded by a broad array of molecular mechanisms, in particular the complex rules for activation of the multiple enhancers that control the expression of pair-rule genes \cite{small+al_91,levine_10}.  One might think that connecting the patterns of gap expression to the positioning of pair-rule stripes would require us to make a model of these mechanisms.  We have taken an alternative view, in which we assume that the embryo makes optimal use of the available information.  Once we make this hypothesis, decoding becomes a well defined mathematical problem, with no free parameters.    The result is a dictionary that translates the expression levels of gap genes into inferred positions, and applying this dictionary to the expression profiles from a single embryo generates a map of inferred position vs actual position.  For wild-type embryos, once we include all four gap genes this map is precise and nearly unambiguous, indicating positions along the anterior--posterior axis with an accuracy of $\sim 1\%$, sufficient to distinguish each cell from its neighbors \cite{dubuis+al_13b}.  For mutant embryos, we find maps that are distorted, imprecise, and in some cases ambiguous.  Using these maps to predict the position of pair-rule expression stripes gives results that are in excellent agreement with experiment, both qualitatively and quantitatively.  The strongly supports the hypothesis of optimality on which our decoding scheme is based.

We draw attention to several aspects of the connection between theory and experiment.  First, we decode positions based on graded expression levels of the gap genes \cite{gaul+jackle_89}.  If we think of the gap genes as forming expression domains that are on or off, then maps are ambiguous even in wild-type embryos (Fig~\ref{SF11}), and we certainly could not make meaningful predictions for stripe positions in the mutants.  Second, we analyze gene expression in absolute units;  the only normalization is global,  which is equivalent to choosing units in which to measure concentration. Contrary to this approach, it often is thought that  biological functions must be robust against presumably uncontrollable variations in absolute concentration.  In fact we find that significant components of the difference, for example, between the wild-type embryo and the {\em tsl} mutant are $10-20\%$ differences in the absolute concentrations of gap gene products  (Fig \ref{SF10}), and these differences propagate through our decoding to generate distorted maps that correctly predict the shifts in pair-rule stripes:  quantitative variations in absolute expression have precise functional consequences.  

Third, we find that maps in the mutants are sufficiently variable that they predict variability in the {\em number} of pair-rule gene stripes, not just their locations; such stripe number variations essentially never occur in wild-type embryos.  One might have thought that reproducibility of stripe numbers emerges only as a  property of the network of interactions among the different pair-rule genes, and even among neighboring cells. Instead, we see that stripe numbers can be reproducible in wild-type embryos because the positional signal carried by gap gene expression levels is extremely precise, with so little noise that a direct local readout of these signals generates near zero probability of dropping or adding a stripe.  Deleting one or two of the maternal inputs to the gap gene network pushes the system into a regime where noise levels are higher, and applying our dictionary to these noisier expression profiles correctly predicts the occurrence of stripe variability.  This is consistent with the idea that the naturally occurring inputs are matched to the signal and noise characteristics of the network \cite{tkacik+al_08a,dubuis+al_13b}.

In many ways our maps of implied position as a function of actual position provide a quantitative, probabilistic version of the older  idea that one can plot cell fate vs position---a fate map---even in mutants; see, for example, Ref  \cite{schupbach+wieschaus_86}.  In its original form, this notion of a map depends on the fact that   what we see in the mutant are rearrangements, deletions, and duplications, but no new pattern elements.  It usually is assumed that this arises from canalization \cite{waddington_42,waddington_59}: although the early stages of pattern formation must generate new and different signals in response to the mutation, subsequent stages of processing force these signals back into a limited set of possibilities set.  What we see here is that even rather early signals, those which are responding immediately to the primary maternal inputs, can be decoded to recapitulate the patterns seen in the wild-type. There is no need for subsequent steps to drive the pattern back to something built from wild-type elements, since it already  is  in this form.

The theoretical framework we have developed here makes additional, testable predictions.  Simultaneous measurements of pair-rule expression with all of the gap genes would allow us to test directly whether, for example, the predicted variations in stripe number are correct, embryo by embryo, rather than just in aggregate.  More subtly, since there are spatial correlations in the fluctuations of gap gene expression levels \cite{krotov+al_14}, our decoding predicts that there should be correlations in the small positional errors that occur even in wild-type embryos, and hence the fluctuations in position of the pair-rule stripes must also be correlated.    These correlations should be different in the mutants, in ways that can be predicted quantitatively.  Most fundamentally, the molecular mechanisms that lead from gap gene product concentrations to pair-rule expression must, effectively, implement the dictionary that we have developed. Thus, we should be able to predict the functional logic of these developmental enhancers by asking that they provide an optimal decoding of positional information, rather than fitting to data.  

Perhaps the most important qualitative conclusion from our results is that precision matters.  We are struck by the ability of embryo to generate a body plan that is reproducible on the scale of single cells, corresponding to positional variations $\sim 1\%$ of the length of the egg.  As with other examples of extreme precision in biological function, from molecule counting in bacterial chemotaxis to photon counting in human vision, we suspect that this developmental precision is a fundamental observation, and to the extent that precision approaches basic physical limits it can even provide the starting point for a theory of how the system works \cite{bialek_12}.  But precision in the final result of development could arise from many paths.  We have a theoretical framework that suggests how such precision could arise from the very earliest stages in the control of gene expression, if this control itself is very precise, and this has motivated experiments to measure gene expression levels with correspondingly high experimental precision.   What we have done here is to bring theory and experiment together, providing a parameter-free prediction of how quantitative variations in gap gene expression levels should influence the developmental process on the hypothesis that the embryo makes optimal use of the available information, in effect maximizing precision at every step.  Genetics then gives us a powerful tool to test these predictions, manipulating  maternal inputs and observing pair-rule outputs.  These rich data are in detailed agreement with theory, providing strong support for the precisionist view.

\begin{acknowledgments}
We thank JO Dubuis and R Samanta for help with the experiments, and M Biggin and N Patel for sharing antibodies used in the pair-rule gene measurements. This work was supported, in part, by US National Science Foundation Grants PHY--1305525 and CCF--0939370; by National Institutes of Health Grants P50GM071508, R01GM077599, and R01GM097275, by  Austrian Science Fund grant FWF P28844, and by   HHMI (to EFW and through an International Fellowship to MDP).
\end{acknowledgments}


 \appendix
\renewcommand\thefigure{S\arabic{figure}}  
\renewcommand\theequation{S\arabic{equation}}
\setcounter{figure}{0}

\section{Experimental methods}
\label{app_exp}

\begin{figure*}
\centerline{\includegraphics[width =183mm]{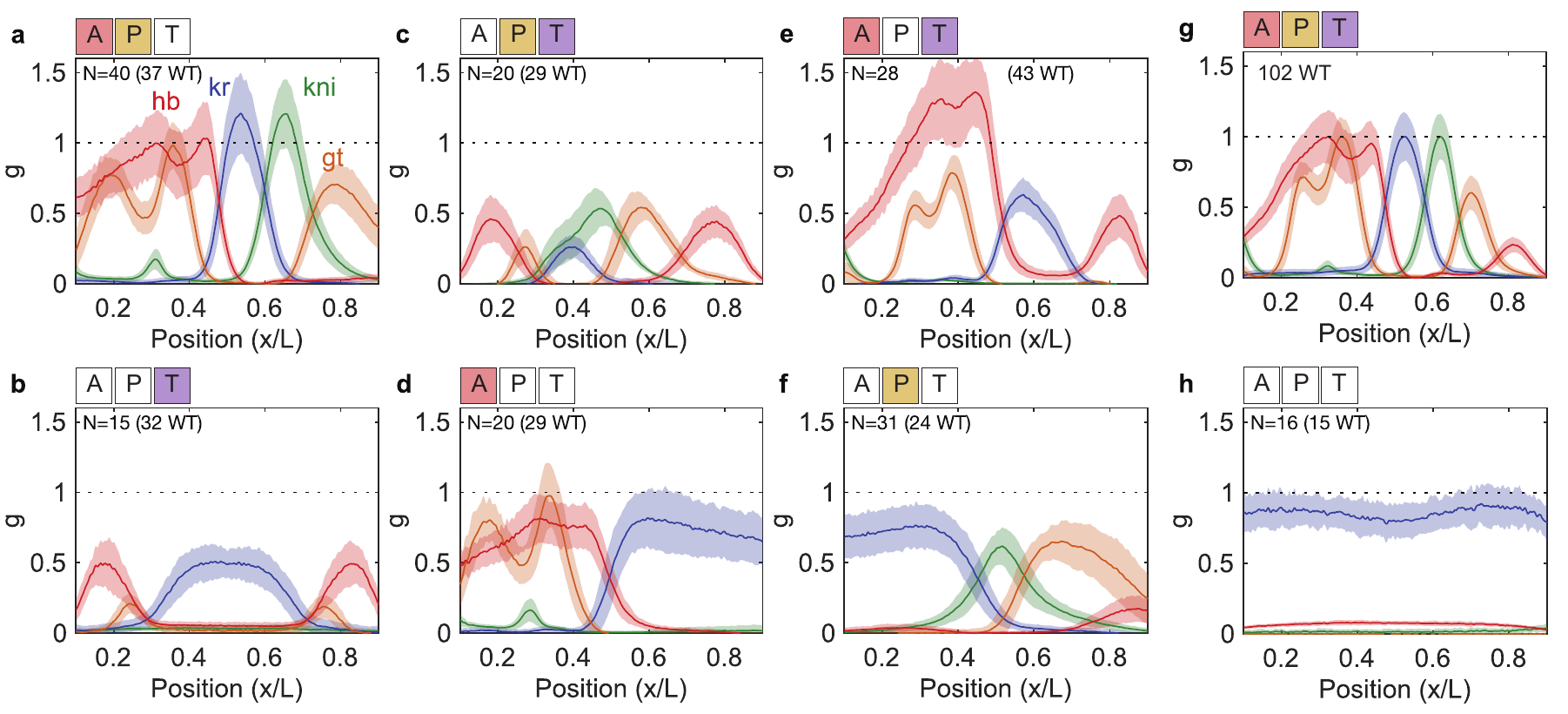}}
\hspace{1cm}
\caption{{\bf Gap gene expression in mutant embryos.} Dorsal gap gene expression profiles (mean $\pm$ SD across embryos aged 38$\!\--\!$48 min into nuclear cycle 14; N indicates number of embryos) in mutant backgrounds. The expression levels $g$ are measured in units of maximal wild-type expression levels, which are measured from on-slide wild-type embryos collected in the same time window (number of embryos is shown in parenthesis).  {\bf a-b.} Terminal system (via {\em tsl}),  {\bf c-d.} Anterior system (via {\em bcd}), {\bf e-f.} Posterior system (via {\em nos}), is absent or the only input of positional information. For completeness we also show the gap gene expression profiles in ({\bf g.}) wild-type, and ({\bf h.}) triply mutant embryos. \label{SF1}}  
\end{figure*}

\leftline{\em Fly strains} 
\smallskip

Embryos lacking single maternal patterning systems were obtained from females homozygous for $bcd^{E1}$, $osk^{166}$ or $tsl^4$.   For embryos with positional information only from the Osk patterning system, we used females homozygous for $bcd^{E1}$ $tsl^4$.   For embryos with only information from Bcd, we used {\it y w} p[{\it w+ Bcd-GFP}]; $bcd^{E2}$ $osk^{166}$ $tsl^4$  females.   p[{\it w+ Bcd-GFP}] encodes a fully rescuing GFP tagged Bcd protein~\cite{gregor+al_07b}.  To obtain embryos with input only from the Torso patterning system, we used $bcd^{E2}$ $osk^{166}$ females for gap gene measurements and $bcd^{E1}$ $nos^{BN}$ females for pair-rule embryos.   The segmentation phenotypes of $osk^{166 }$and $nos^{BN}$ are equivalent~\cite{wang+al_94}.    Embryos lacking all maternal patterning systems were obtained from triply mutant $bcd^{E1}$ $nos^{BN}$ $tsl^4$  females.  All stocks were balanced with TM3, Sb.

\smallskip
\leftline{\em Measuring gap gene expression}
\smallskip

Gap protein levels were measured as described by Dubuis et al.~\cite{dubuis+al_13a}. To quantify mutant gap protein levels in units of wild-type protein levels, mutants and wild-type embryos were stained together, and imaged alongside on the same microscope slide in a single acquisition cycle. We draw attention to the discussion of experimental errors in Ref ~\cite{dubuis+al_13a}, because this  is especially important for our analysis.   As described there, we focus on a narrow time window, 38$\!\--\!$48 min into nuclear cycle 14.

Expression levels were normalized such that the mean expression levels of wild-type embryos ranged between 0 (assigned to the minimal value across the AP axis of the mean spatial profile, separately for each gap gene) and 1 (similarly assigned to the maximal value across the AP axis). In detail, gene expression profile $g_{\rm i}^{\alpha}$ of any embryo $\alpha$ was calculated as: 
\begin{equation}
g_{\rm i}^{\alpha} =
\frac{I^{\alpha}_{g_{\rm i}}- \bar I^{\text{wt}}_{\text{min},g_{\rm i}}}
{\bar I^\text{wt}_{\text{max},g_{\rm i}}-\bar I^{\text{wt}}_{\text{min},g_{\rm i}}}
\label{normeq}
\end{equation}
where $\bar I^{\text{wt}}_\text{min}$ and $\bar I^{\text{wt}}_\text{max}$ are the lowest and highest raw fluorescence intensity values of the mean wild-type embryo fluorescence profiles; $I^\alpha_{g_i}$ is the raw fluorescence profile of the particular embryo, which can be either mutant or wild-type.  Note that this normalization simply assigns a conventional unit of measurement to gap gene concentrations; no per-embryo profile ``alignment'' is used to  reduce embryo-to-embryo variance. Furthermore, mutant embryos were normalized to their wild-type reference for each gap gene, so absolute changes in gap gene concentrations---not only changes in the shape of the gap gene spatial profiles---were retained in all analyses. A summary of results on the mutant gap gene expression profiles (mean $\pm $ SD across embryos) is given in Fig~\ref{SF1}.

\begin{figure*}
\centerline{\includegraphics[width =.85\linewidth]{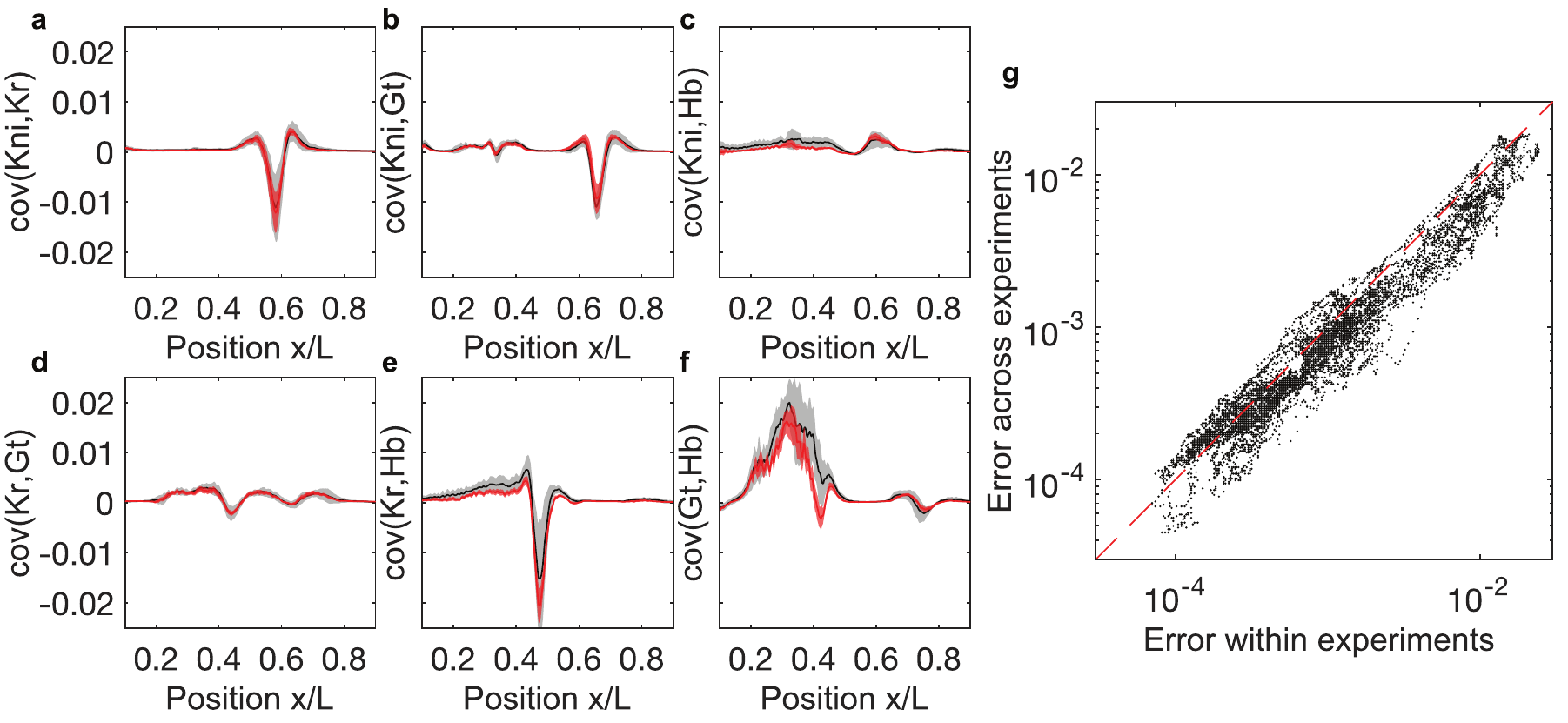}}
\caption{{\bf Estimation of gap gene covariance matrix from wild-type embryos}. For each of $7$ wild-type datasets ($N =37,29,43,32,29,24$, and $102$ embryos) we compute the covariance matrix of fluctuations in gap gene expression levels at each point along the AP axis,  Errors within an experiment are standard deviations across matrices computed from random halves of the data, while errors across experiments are the standard deviations of the $7$ means.   {\bf a-f.} Off-diagonal covariance matrix elements at each point along the AP axis;  mean (black) $\pm$ errors across experiments (grey shading). For reference, we also show the covariance matrix elements from the single largest wild-type dataset ($N=102$) embryos (red) and the errors within this experiment (red shading). {\bf g.} Scatter plot of error within experiment (largest value from the $7$ datasets) vs error across experiment on estimating all covariance matrix elements.   \label{SF12}}
\end{figure*}

 \smallskip
\leftline{\em Measuring pair-rule gene expression}
\smallskip

To image pair-rule proteins, we used guinea pig anti-Runt, and rabbit anti-Eve (gift from Mark Biggin) polyclonal antibodies, and monoclonal mouse anti-Pax3/7(DP312) antibody (gift from Nipam Patel). Secondary antibodies are, respectively, conjugated with  Alexa-594 (guinea pig), Alexa-568 (rabbit), and Alexa-647 (mouse) from Invitrogen, Grand Island, NY.  Embryo fixation, antibody staining, imaging and profile extraction were performed as described by Dubuis et al.~\cite{dubuis+al_13a}.

Our goal was to predict features of pair-rule protein concentration profiles, such as the locations of expression peaks, for which comparisons between wild-type and mutant expression levels of pair-rule genes were not essential.  Pair-rule protein profiles were measured in mutant embryos in time widows of 45- to 55-min into nuclear cycle 14; for consistency with gap gene analyses and convenience we normalized such that the mean expression levels for each gene in each batch of embryos ranged between 0 and 1; individual profiles were scaled to minimize $\chi^2$, as in Refs~\cite{gregor+al_07b, dubuis+al_13a}.  As an exception, we report pair-rule expression levels in triple maternal mutants ({\em bcd nos tsl}) in wild-type units, because the pair-rule genes are expressed uniformly and therefore lack positional features.

\begin{figure*}[t]
\centerline{\includegraphics[width = .9\linewidth]{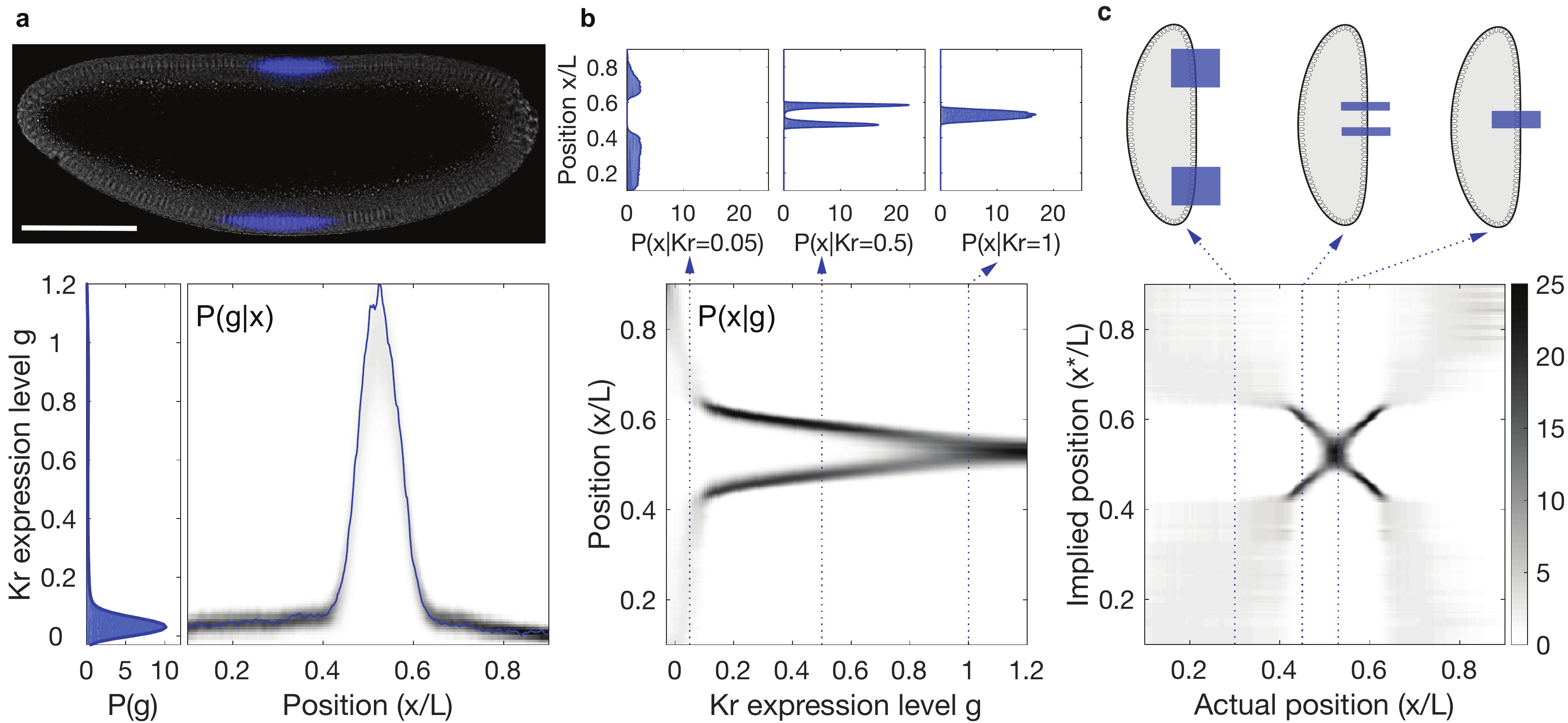}}
\caption{{\bf Coding and decoding of position in the fly embryo based on expression of a single gap gene.} {\bf a.} Optical section through the midsagittal plane of a {\em Drosophila} embryo with immunofluorescence labelling for  {Kr\"uppel} (Kr) protein. Raw dorsal fluorescence intensity profile of depicted embryo (blue) and encoding probability distribution $P(Kr|x)$ (gray) constructed from 38 wild-type embryos of ages between 40--44 min into nuclear cycle 14. Position $x$ along the anterior-posterior axis is normalized by the length $L$ of the embryo; $x/L=0$ corresponds to the anterior end of the embryo, and $x/L=1$ corresponds to the posterior end. Probability distribution of Kr expression levels (right). {\bf b.} Decoding probability distribution $P(x|Kr)$ constructed via Bayes' rule from the measured probability distributions $P(g)$ and $P(g|x)$ in {\bf a}, using a uniform prior $P(x)=1/L$.  The distribution $P(x|Kr)$ is the optimal decoder, which maps Kr levels to positions along the AP axis. For example, the probability distributions of locations $x$ consistent with observing Kr levels  0.05, 0.5, or 1 (main), are the conditional probability densities $P(x|Kr)$ shown in the three top panels.  {\bf c.} Decoding map $P_g^\alpha (x^*|x )$ for a single embryo (depicted in {\bf a}),  where $\alpha$ is the embryo index, running from 1 through 38. For three locations, cartoons (top) display uncertainties and ambiguities in determining location in the embryo baed on Kr alone.  Importantly, only single-gene decoders (e.g. the distribution $P(x|Kr)$ in {\bf b}) can be directly visualized (decoding with two genes, for instance, requires a 4-dimensional visual representation). Decoding maps $P(x^*|x)$, however, can be visualized for an arbitrary number of genes.  \label{SF0}}  
 \end{figure*}

 \section{Theoretical methods}
\label{methods+thy}

To construct decoding maps and subsequently predict pair-rule expression stripes, Eqs (\ref{decode1gene}) and (\ref{map1}) require us to estimate the distribution of gap gene expression levels at each position, $P(\{g_{\rm i}\}|x)$, from data. Direct sampling is unfeasible even for just a single gene. Instead, we approximated the embryo-to-embryo fluctuations in gene expression as Gaussian with mean and (co)variance that vary with position.   In previous work we tested this approximation; while we can see deviations from Gaussianity \cite{krotov+al_14}, the Gaussian approximation  gives very accurate estimates of the positional information carried by the expression levels of individual genes \cite{dubuis+al_13b,tkacik+al_15}, which is most relevant for the decoding that we attempt here.
 
The Gaussian approximation for a single gene corresponds to
 \begin{equation}
P(g|x) = {1\over{\sqrt{2\pi\sigma_g^2(x)}}} e^{-\chi_1^2(g,x)/2} , \label{s1gauss}
\end{equation} 
where $\chi^2_1(g,x)$ measures the similarity of the gene expression level to the mean, $\bar g (x)$,  at position $x$, 
\begin{equation}
\chi^2_1\left( g  , x \right) = {{(g - \bar g (x))^2}\over{\sigma_g^2 (x)}} , \label{s2gauss}
\end{equation}
and $\sigma_g (x)$ is the standard deviation in expression levels at point $x$ (e.g., shown by the shading in Fig \ref{SF1}).  Given measurements of gene expression vs position in a large set of embryos, we can compute the mean and variance in the standard way, so that Eqs~(\ref{s1gauss}-\ref{s2gauss}) can be applied directly to the data.

The generalization of the Gaussian approximation to the case where coding and decoding are based on a combination of $K$ genes simultaneously is given by
\begin{equation}
P (\{g_{\rm i}\} | x  ) 
=
 \frac{1}{\sqrt{(2\pi)^K\det[C(x)]}}e^{- \chi_K^2\left( \{g_{\rm i} \}, x \right)/2}
\label{mapK}
\end{equation}
where  $\chi_K^2$ measures the similarity of the gene expression pattern  to the mean {\em pattern} expected at $x$,
\begin{equation}
\chi_K^2\left(\{g_{\rm i}\} , x \right) = 
\sum_{{\rm i}, {\rm j} = 1}^K \left( g_{\rm i} - {\bar g}_{\rm i}(x)\right) \left( C^{-1} (x)\right)_{\rm ij} 
 \left( g_{\rm j} - {\bar g}_{\rm j}(x)\right) ,
 \label{chisq}
\end{equation}
and $C(x)$ is the covariance matrix of fluctuations in the expression of the different genes at point $x$.  Figures \ref{SF12}a-f shows the estimation of $4\times 4$ covariance matrix of gap gene fluctuations across embryos, $C_{\rm ij} = \langle (g_{\rm i}^{\alpha} - \langle g_{\rm i}^{\alpha}\rangle)(g_{\rm i}^{\alpha}-\langle g_{\rm i}^{\alpha}\rangle)\rangle$ (where $\langle \cdot \rangle$ denotes averaging over embryos $\alpha$). Note that the covariance matrix, as well as the mean profiles themselves, are a function of position along the AP axis.   Comparing the covariance matrix estimates across replicates of wild-type datasets, Fig~\ref{SF12}g  shows that our data are sufficient for us to have control over estimation errors, so that Eqs~(\ref{mapK}-\ref{chisq}) can likewise be applied directly to the data.

\section{The decoding dictionary}

\begin{figure}
\centerline{\includegraphics[width =  \linewidth]{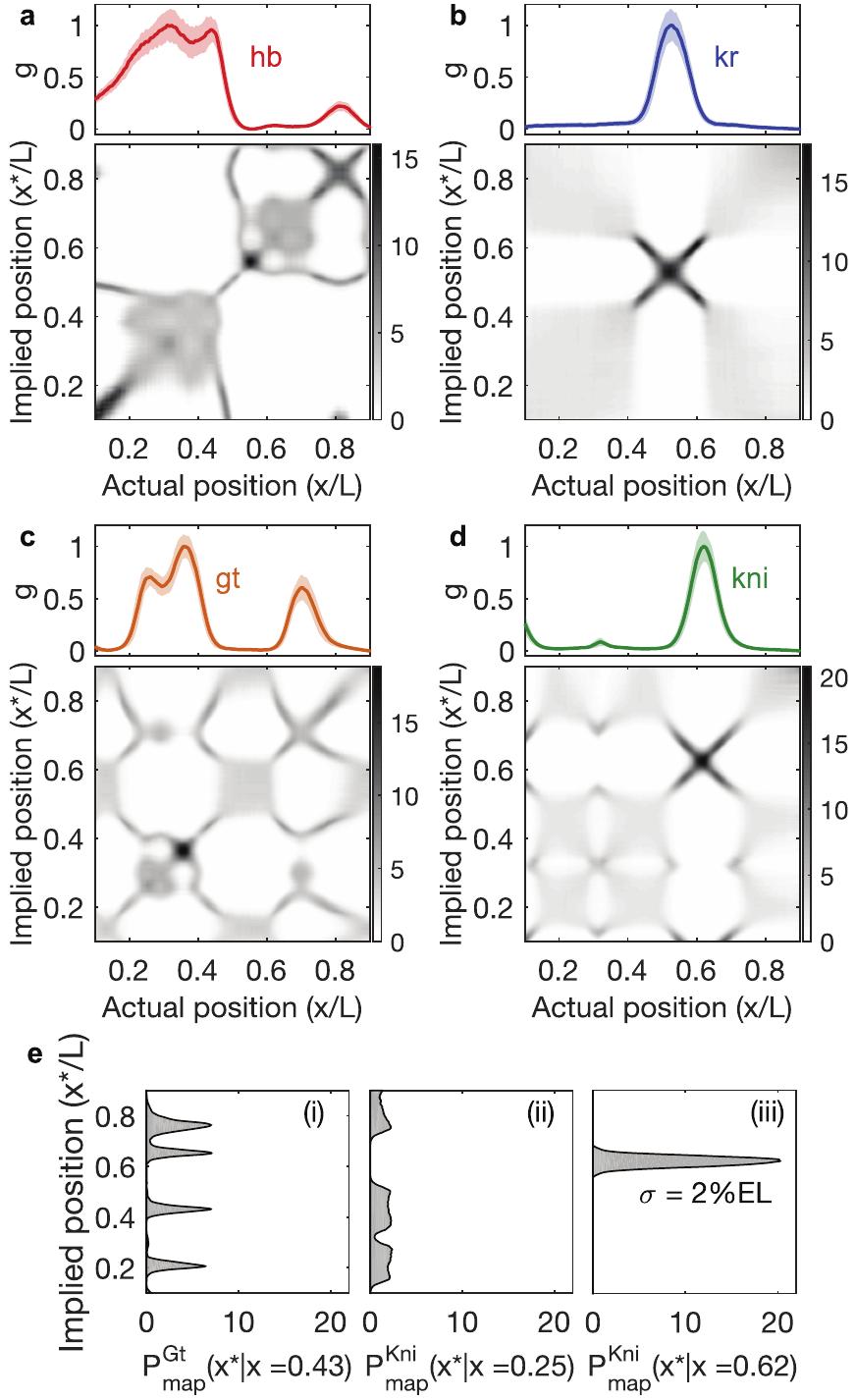}}
\caption{{\bf Decoding maps from four individual gap genes.} {\bf a.--d.} Top row: dorsal expression profile (mean $\pm$ SD across 38 wild-type embryos of ages 40$\!\--\!$44 min into n.c. 14); gene as indicated in panel. Bottom row: average  decoding maps constructed by decoding from genes profiles in panel above. {\bf e.}  Three sample features of the decoding maps - (i) precise, but ambiguous, (left), (ii) imprecise and ambiguous (middle), and (iii) unambiguous and  precise (right). The latter probability density is Gaussian distributed with positional error $\sigma_x=2\%$ of the embryo length (EL).  The Kr map in {\bf b} is as in Fig\ref{F2}a, it is obtained by averaging all decoding maps from individual (see Fig\ref{SF0}c) embryos. \label{SF2}}
\end{figure}

\begin{figure}[t]
\centerline{\includegraphics[width = \linewidth]{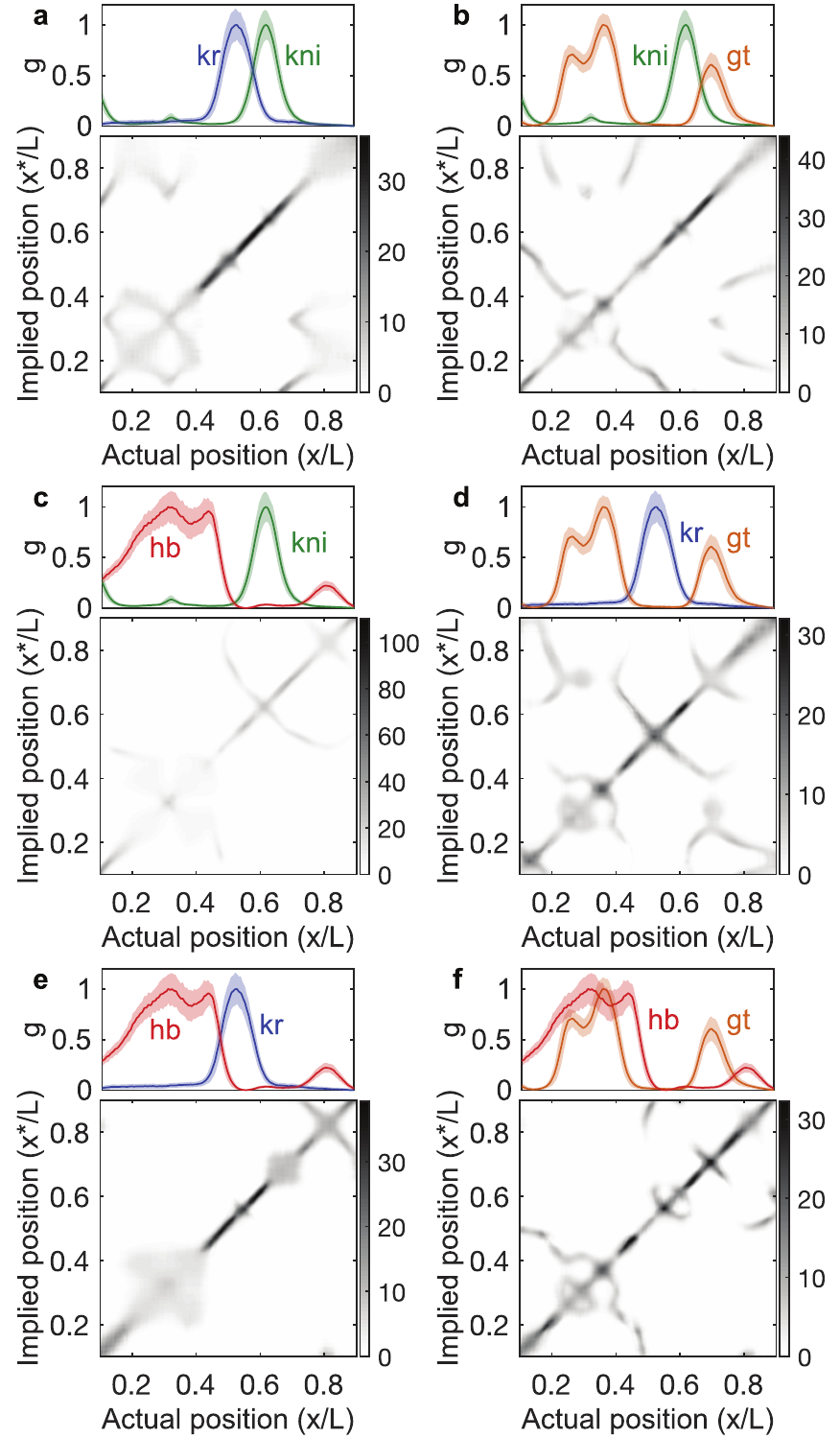}}
\caption{{\bf Decoding based on pairs of gap gene expression profiles.} {\bf a.--d.}  Top: dorsal gap gene expression profiles (mean $\pm$ SD as in Fig.\ref{SF2}). Bottom:  average decoding maps constructed by decoding from pairs of gap genes indicated in top panels. The Hb--Kr combination in {\bf e} is also Fig \ref{F2}b.  \label{SF3}}
\end{figure}

\begin{figure}
\centerline{\includegraphics[width = \linewidth]{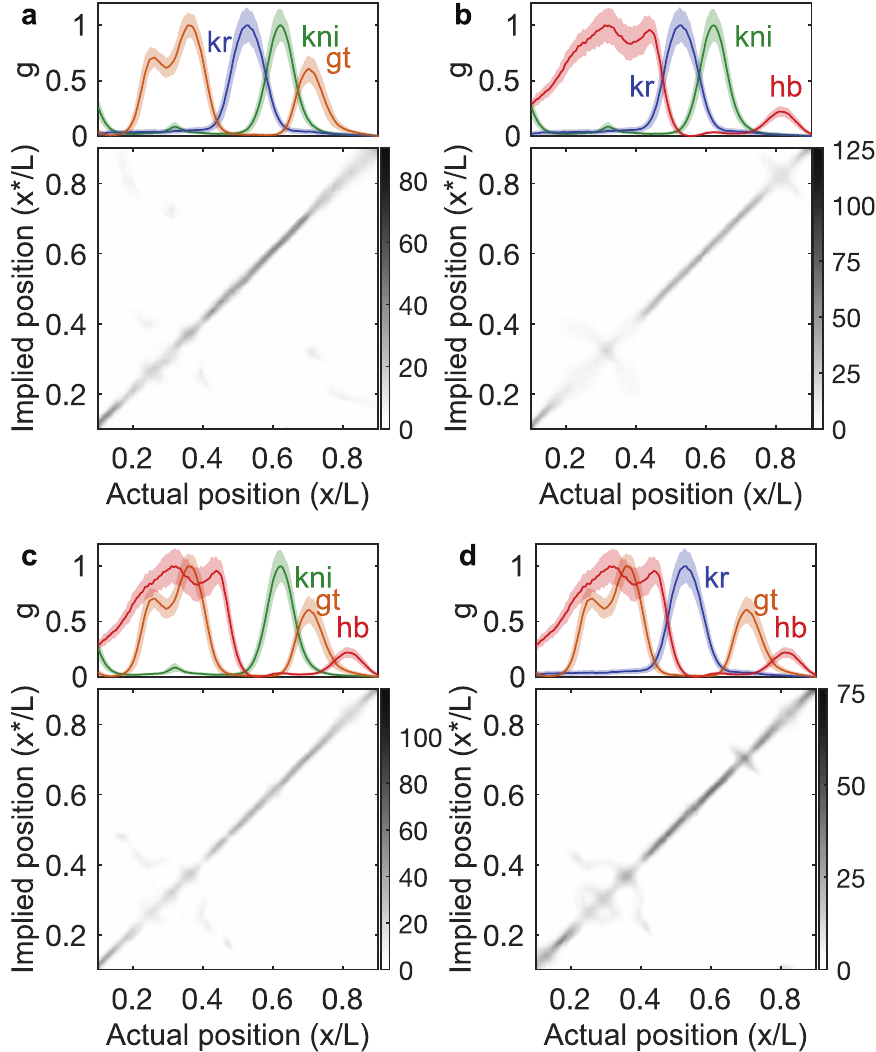}}
\caption{{\bf Decoding based on sets of three gap gene expression profiles. } {\bf a.--d.} Top: dorsal gap gene expression profiles (mean $\pm$ SD as in Fig.\ref{SF2}). Bottom:  average decoding maps constructed by decoding from sets of three gap genes as indicated in top panels;  {\bf d} is identical to Fig \ref{F2}c. \label{SF4}}
\end{figure}

Figure~\ref{SF0} shows a step-by-step procedure for constructing a  ``decoding dictionary" based on a single gap gene, Kr, from measured data, and a ``decoding map" for a single wild-type embryo; the decoding map in  Fig~\ref{F2}a is an average over 38 such individual decoding maps. Similarly, top panels of Fig \ref{SF2}a-d show the profiles of all four individual gap genes in the wild-type embryos, while the bottom panels show the corresponding decoding maps.  As with the case of Kr\"uppel in Fig \ref{SF0}, all of these maps show substantial ambiguities, where the signal at one point in the embryo is consistent with a wide range of possible positions.  Ambiguity arises whenever a vertical slice through these density plots encounters multiple peaks, but in the case of decoding based on single genes these ambiguities are so common that they result in either vast swaths of grey  or in intricate folded patterns.  In particular locations---specifically, at the flanks of mean expression profiles where the slope of the profile is high---the distributions $P(x^*|x)$ become highly concentrated, indicating that the quantitative expression levels of individual genes provide the ingredients for precise inferences of position, as suggested in Refs~\cite{gregor+al_07b,dubuis+al_13b}. 

\begin{figure*}
\centerline{\includegraphics[width =\linewidth]{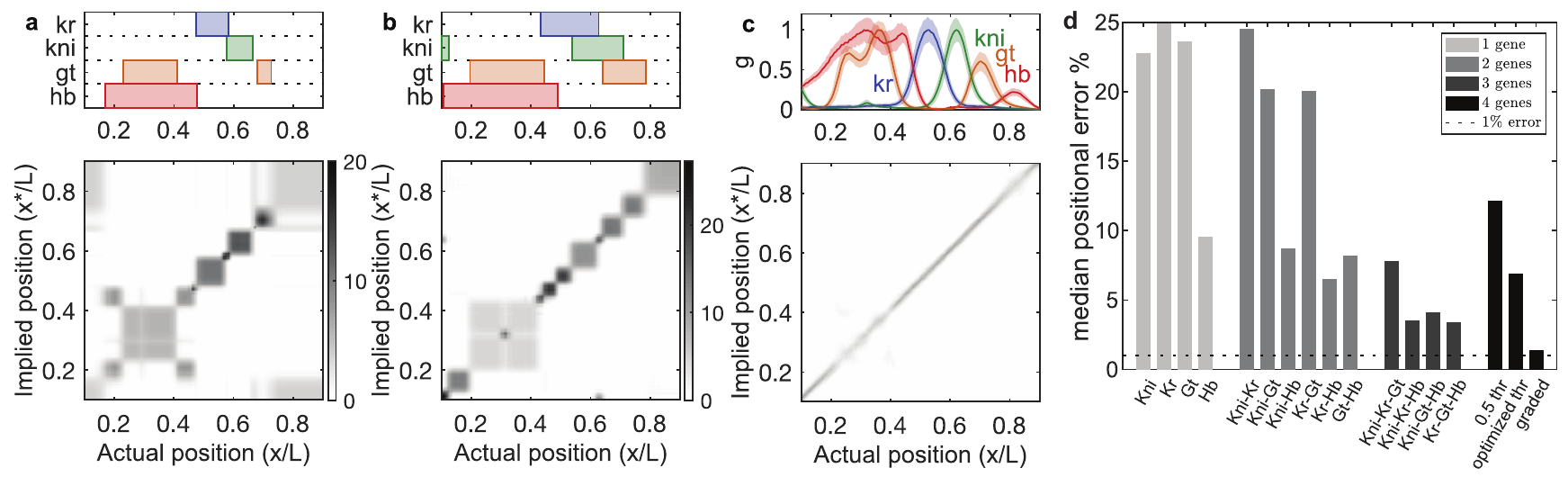}}
\caption{{\bf Decoding based on traditional binary, threshold-based readout is imprecise and ambiguous. a.}  Decoding from gap genes being ON or OFF, with ON state declared when they are expressed at more than half of their maximum mean level (top). {\bf{b}.} As in a, but with thresholds set so that the mutual information between $x^8$ and $x$ is maximized.  {\bf{c}.}  Decoding map based on graded variations in gap gene expression, replica of Figure~\ref{F2}d for comparison.  {\bf d.} Precision of decoding based on different combinations of genes. We compute the standard deviation of the distributions $P(x^*|x)$ and then compute the median over all $x$.  Results are plotted for decoding based on all combinations of 1, 2, and 3 genes, all four genes (``graded''), and  four genes thresholded into on/off. \label{SF11}}
\end{figure*}

Figure~\ref{SF2} shows that combining two genes always reduces ambiguity relative to the single gene case, but does not eliminate it entirely, and a similar trend is observed in Fig \ref{SF3} with triplets of gap genes. The four-gene case shown in Fig~\ref{F2}c sharpens the decoding maps further (cf. the scale of distributions $P(x^*|x)$), achieving a low positional error of $\sim 1.5\%$ across the majority of locations along the AP axis.  We can quantify this sharpening by computing the standard deviation of the distribution $P(x^*|x)$, and then finding the median over $x$; a summary of these results is given in Fig \ref{SF11}d.

Figure~\ref{SF11} further shows that the traditional interpretation of gap genes as generating binary domains of expression separated by sharp boundaries significantly blurs the decoding map, irrespective of whether the gap gene thresholds are selected simply at the midpoint of the expression range (at $g=0.5$) or are adjusted separately for each gap gene to optimize the decoding map. This shows that quantitative, graded  levels of gap gene expression are essential for the precise specification of position.

It is worth emphasizing that the notion of a threshold that determines gene expression (or cell fate) boundaries, which is well defined for a single signal being thresholded, is more ambiguous in the case of multiple concurrent signals that drive patterning, as is the case here with gap genes. The idea of putting independent, and possibly different, thresholds on each of the inputs separately may appear as a natural extension of a single-gene case, but it is important to realize that this idea already entails a drastic (and untested) independence assumption. It would be equally possible that the relevant patterning thresholds act on some unknown, even nonlinear, combination of the four gap gene signals. In particular, in biophysical models of enhancer function where the gene expression is controlled by the concentrations of multiple inputs (and where the threshold is determined by the sigmoid activation function of the enhancer), the interpretation of thresholds applying to nonlinear combinations of inputs is more realistic than the interpretation of different thresholds independently applying to each of the inputs. Furthermore, the picture of independent thresholds acting on individual gap genes leaves completely unanswered the question of how binarized gap gene profiles can be read out in a biophysically realistic fashion to combinatorially drive the expression of their target genes.

\section{Exploring the mutants}

We analyzed patterns of gap gene expression in six mutant lines of flies, deficient in one or two of the three maternal inputs to the gap gene network, as summarized in Fig~\ref{SF1}. To construct decoding maps for the mutants, as in Fig~\ref{F3}, we first computed posterior distributions $P(x|\{g_{\rm i}\})$ as prescribed by Eq~(\ref{decode1gene}) from wild-type embryo data, and evaluated these distributions at gap gene expression levels measured in mutant embryos. But the wild-type expression levels fill only a very small region of the full four dimensional space of possibilities; if the expression levels in mutant embryos fell largely outside this region, then we would be extrapolating too far from the wild-type measurements and could not make reliable inferences. To test whether this could be the case, we computed $\chi^2$  [Eq (\ref{chisq})] between the observed combinations of expression levels and the mean expression levels expected at each position in the wild-type, and compared that to the $\chi^2$ values for mutant embryos. 

\begin{figure*}
\centerline{\includegraphics[width = \linewidth]{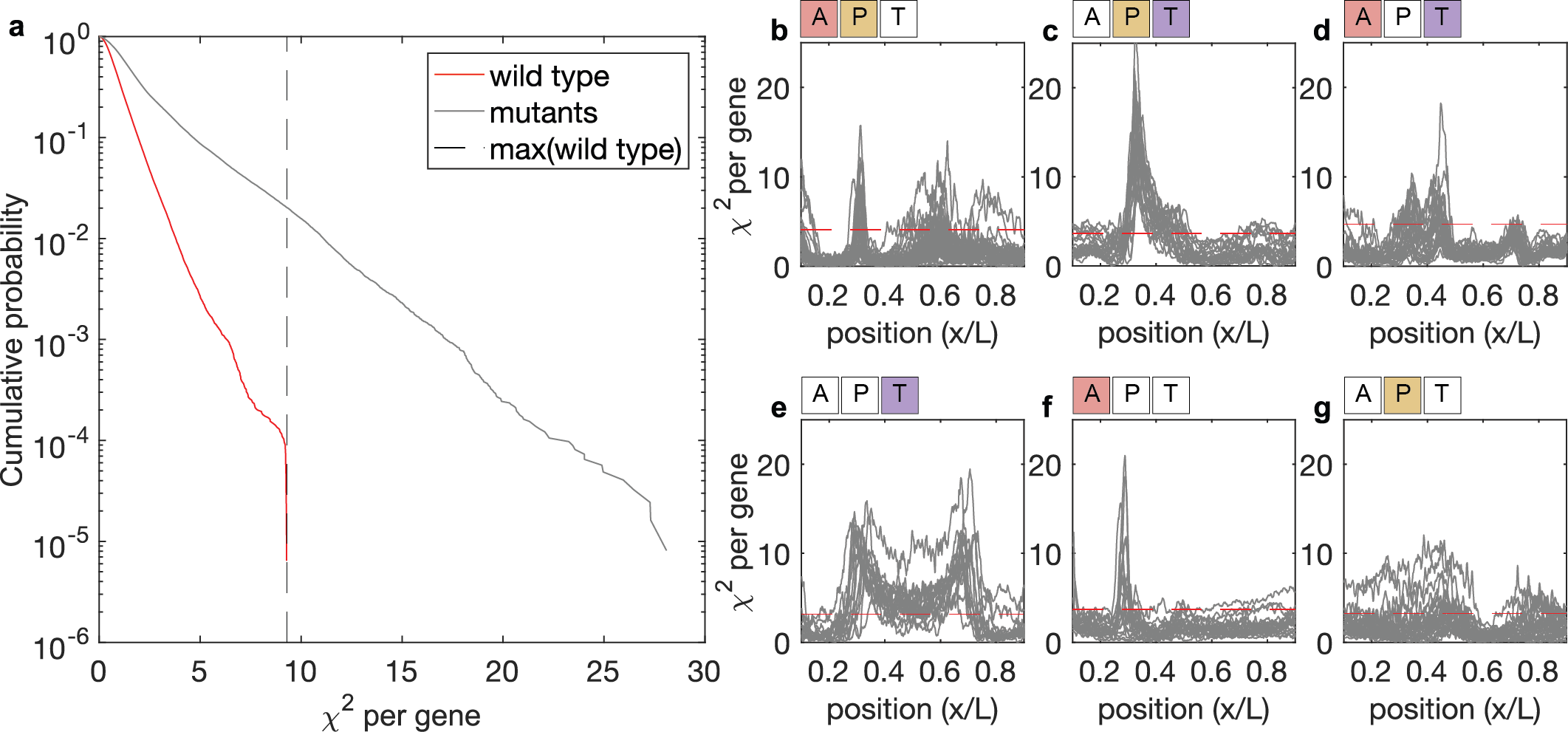}}
\caption{{\bf Gap gene expression levels in mutants largely overlap   those observed in wild-type embryos.}   Normalized sqaured deviation $\chi^2$ between observed and mean expression of all four gap genes.  As explained in the text, we compute this directly for wild-type embryos, while for mutant embryos we compute the minimum $\chi^2$ over all possible correspondences to wild-type positions.   {\bf a.} Cumulative probability (y-axis, log scale) as a function of $\chi^2$ per gene---$\chi^2$ from Eq~(\ref{chisq}), divided by 4. It represents the probability that $\chi^2$ per gene is greater than the value on the x-axis in wild-type embryos (red), and mutant embryos (black).  Vertical dashed line marks the maximal $\chi^2$ observed in wild-type data set;  the  intersection of dashed line with black line shows that this variation in wild-type encompasses $98\%$ of the points in mutants.    {\bf b-g.}  $\chi^2$ per gene for individual mutant embryos as a function of position along the AP axis (grey lines), together with (more conservative) limits on the largest $\chi^2$ per gene observed in wild-type embryos in that particular batch ($1\%$ of the wild-type embryos have $\chi^2$ per gene values larger than denoted by the horizontal red dashed lines); we observe values of  $\chi^2$ per gene which are below the red dashed lines in $\sim90\%$ of the positions,. \label{SF5}}
\end{figure*}

Figure~\ref{SF5} shows the cumulative distribution of $\chi^2$ across the entire population of wild-type embryos, from all six experiments.  Normalized per gene, the mean of $\chi^2$ is one, but the distribution has a tail extending to $\sim10\times$ this value.  To construct a comparable distribution for the mutants, we first note that the gene expression values at one point $x$ in the embryo can be decoded to a position $x'$ that is very far from $x$, as shown by Fig~\ref{F3} and explained in the corresponding discussion in the main text. Consequently, in the mutants we looked for the point $x'$ in the wild-type that achieved the minimum of $\chi^2_K(\{g_{\rm i}\},x')$ over all possible $x'$ (which is the location that the mutant gap gene profiles decode to) and then look at the cumulative distribution of $\chi^2$ at these decoded locations. 

As expected, $\chi^2$ values in the mutant are larger than in the wild-type, but there is a surprising degree of overlap between the two distributions:  the largest value of $\chi^2$ that we observe in the wild-type embryos is larger than $98\%$ of the values that we see in the mutants. Although mutant background induces huge changes in the inputs of the gap gene network and in the gap gene profiles themselves, the gap gene network responds in a way that is not so far outside the distribution of possible responses under natural conditions.  This fact is  what makes decoding  positional information in the mutants feasible.

\section{Predicting stripe positions}
\label{app:stripes}

Decoding maps make parameter--free predictions for the locations of positional markers in  mutant embryos.  To test these predictions, we compare  to the  locations of expression peaks for the pair-rule genes.
If a cell at position $x$ in the mutant embryo has expression levels for the gap genes that lead to a high probability of inferring a position $x^* = x_s$, where $x_s$ is the position of a pair-rule stripe in the wild-type, then we expect that there will be a peak in pair-rule gene expression at the point $x$ in the mutant.  Mathematically, this process (shown graphically in Fig \ref{F3}) proceeds as follows: we construct $P^\alpha_{\rm map}(x^*|x)$ and look at the line $x^* = x_s$; this gives us a (non--normalized) density $\rho_s^\alpha(x) = P^\alpha_{\rm map}(x^* = x_s|x)$, and there should be pair-rule stripes at the local maxima of this density.  Because stripes in the wild-type are driven by different enhancers and are thus not identical, it is important that our calculation should predict the occurrence of a particular identified stripe $s$ at  $x$.

\begin{figure}
\centerline{\includegraphics[width = \linewidth]{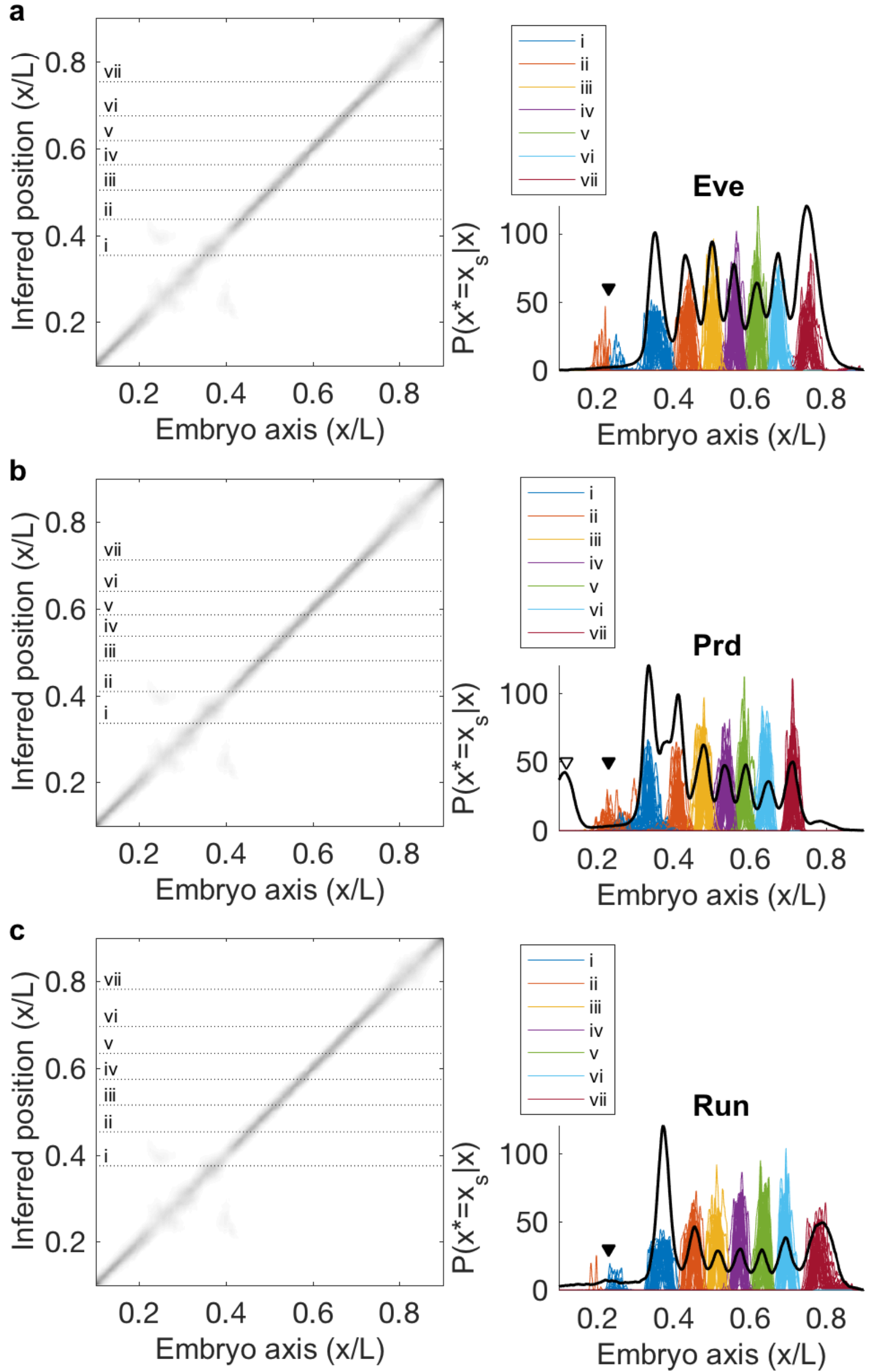}}
\caption{{\bf Predicting pair-rule stripe locations in wild-type embryos.}  We trace horizontal intersections of  probability density $P(x^*|x)$  maps in 38 individual wild-type embryos at the average locations of pair-rule peaks, to yield the densities $\rho_s^{\alpha}(x)$ defined in the text.  Rows are for the genes {\em eve} ({\bf a.}),  {\em Prd} ({\bf b.}), and {\em run} ({\bf c.} ). Left panels: average decoding map (as in Fig.~\ref{F2}d) with horizontal dotted lines marking the average locations of pair-rule peaks; Roman numerals indicate stripe number. Right panels: $\rho_s^{\alpha}(x)$, with colors marking different stripes $s$ (legend). For reference, the average pair-rule expression in wild-type is plotted (black solid line),  scaled for visualization.   We exclude the anterior-most Prd stripe (open triangle) from further analysis, because it is not well defined. \label{SF15}}
\end{figure}

The construction of the density  $\rho_s^\alpha(x)$ is shown in Fig~\ref{SF15} for each stripe of Eve, Prd, and Run, and for each individual wild-type embryo. There is an excellent correspondence between the average pair-rule gene expression profile and the set of individual embryo densities for all stripes; the sole discrepancy appears to be a small ambiguity in the decoding map that hints at two weak ``echoes'' of pair-rule stripes 1 and 2 in the far anterior (for $x<0.3$), which we did not detect in the data. These may be missing because of influences from other gap genes that are active in the far anterior.

\begin{SCfigure*}[1]
\hspace{1.5cm}
\centerline{\includegraphics[width =120mm]{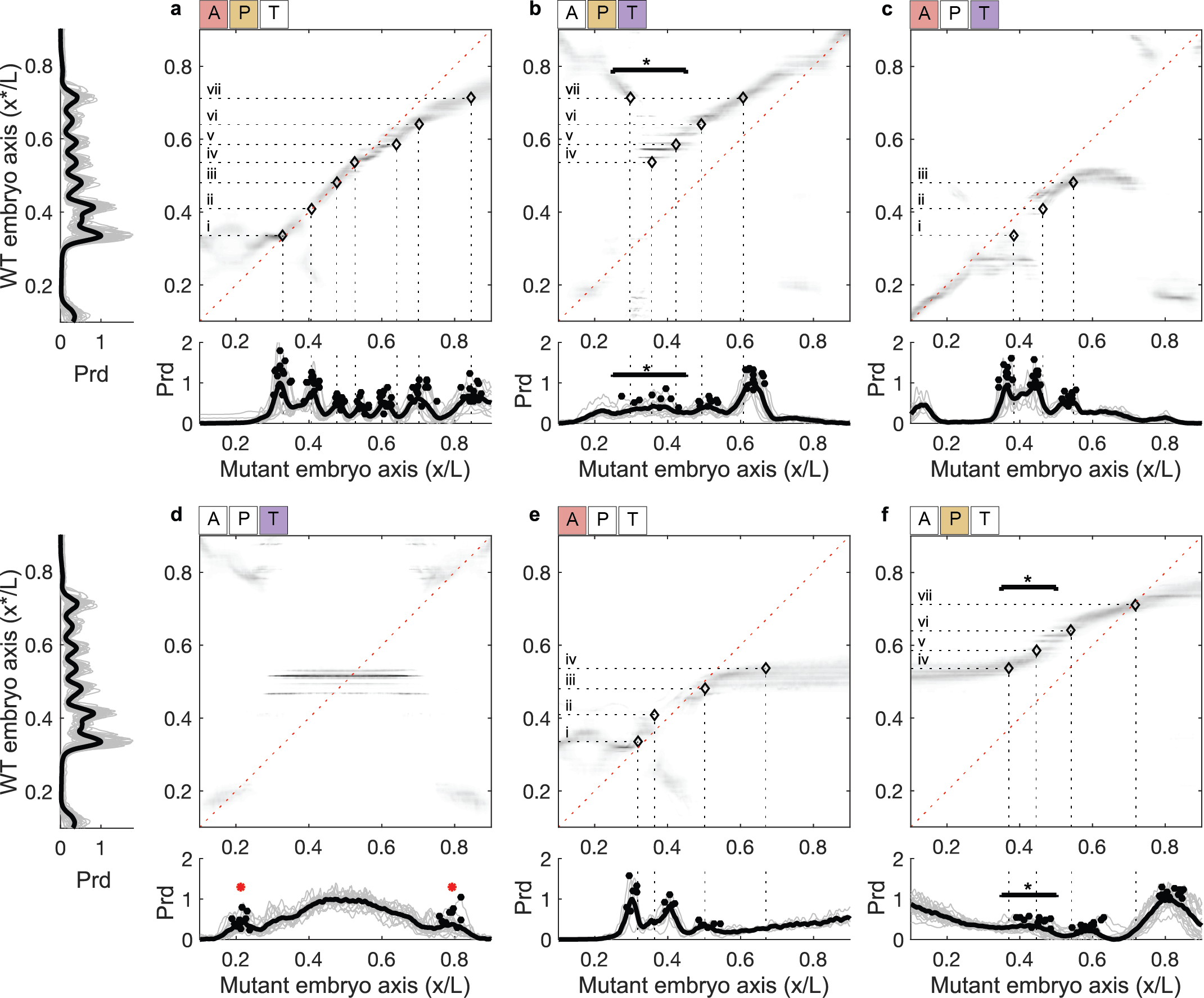}}
\hspace{1.75cm}
\caption{{\bf Decoding from mutant embryos and predicting the locations of Prd stripes.} Average decoding maps for six maternal mutant backgrounds: {\bf a.} {\it etsl};  {\bf b.} {\it bcd$^{\rm E1}$}; {\bf c.}  {\it osk};  {\bf d.} {\it bcd$^{\rm E1} $osk}; {\bf e.} {\it osk tsl}; {\bf f.} {\it bcd$^{\rm E1} $tsl}. In each decoding panel, we use the average locations of the seven peaks of wild-type Prd expression (left side of panels {\bf a} and {\bf d}) to predict Prd stripe locations in the mutant backgrounds: where horizontal dotted lines intersect the probability density (marked by open black circles and vertical dotted lines). Measurements of the actual Prd expression profiles in each mutant background are shown below the corresponding decoding panel, where filled black circles indicate the profile peaks. Intensity in all decoding panels refers to wild-type intensity in Fig 2d. Roman numerals above the horizontal dotted lines denote the wild-type Prd stripe number. Horizontal starred bars (panels {\bf b} and {\bf f}) indicate locations where the expressed number of Prd stripes is variable, which is captured qualitatively by the decoding maps. The red stars in panel {\bf d} mark peaks with variable expressivity, which are not predicted by the decoding map. \label{SF6}}
\end{SCfigure*}

\begin{SCfigure*}
\hspace{1.5cm}
\centerline{\includegraphics[width =120mm]{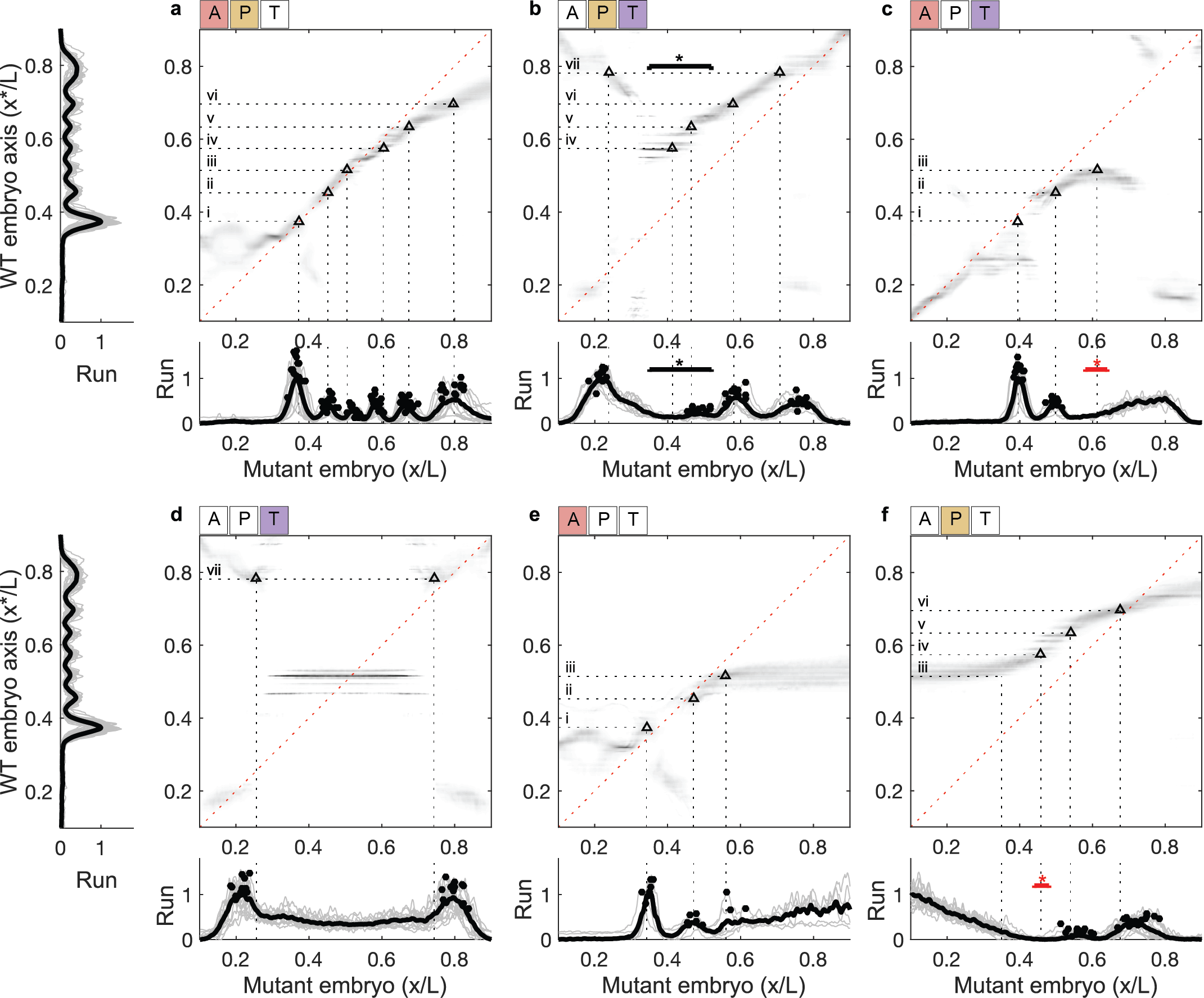}}
\hspace{1.75cm}
\caption{{\bf Decoding from mutant embryos and predicting the locations of Run stripes.} Average decoding maps for six maternal mutant backgrounds: {\bf a.} {\it etsl};  {\bf b.} {\it bcd$^{\rm E1}$}; {\bf c.}  {\it osk};  {\bf d.} {\it bcd$^{\rm E1} $osk}; {\bf e.} {\it osk tsl}; {\bf f.} {\it bcd$^{\rm E1} $tsl}. In each decoding panel, we use the average locations of the seven peaks of wild-type Run expression (left side of panels {\bf a} and {\bf d}) to predict Run stripe locations in the mutant backgrounds: where horizontal dotted lines intersect the probability density (marked by open black circles and vertical dotted lines). Measurements of the actual Run expression profiles in each mutant background are shown below the corresponding decoding panel, where filled black circles indicate the profile peaks. Intensity in all decoding panels refers to wild-type intensity in Fig 2d. Roman numerals above the horizontal dotted lines denote the wild-type Run stripe number. Horizontal starred bars (panels {\bf b} and {\bf f}) indicate locations where the expressed number of Run stripes is variable, which is captured qualitatively by the decoding maps. Horizontal starred bars in panels {\bf c,f} mark predicted peaks, which are not observed. \label{SF7}}
\end{SCfigure*}

For ease of visualization, we often look just at the \emph{average} decoding map, as  in Fig~\ref{F3} as well as  Figs~\ref{SF6} and  \ref{SF7}, below. This average, $P_{\rm map}(x^*|x) = \langle P^\alpha_{\rm map}(x^*|x)\rangle_\alpha$, which can be easily plotted as a single map, and then decoded analogously to the procedure outlined above: we looked for the position $x$ where the decoding map peaks if the inferred position $x^*$ were equal to a known pair-rule stripe location in the wild-type. Decoding the ``mean pair-rule stripe position'' in this manner does not differ from decoding single embryos to predict the pair-rule stripe positions individually, and then taking the average prediction, but now we can also predict fluctuations in stripe locations, a fact we used in making Fig~\ref{F4}.

\begin{SCfigure*}
\hspace{1.5cm}
\centerline{\includegraphics[width =110mm]{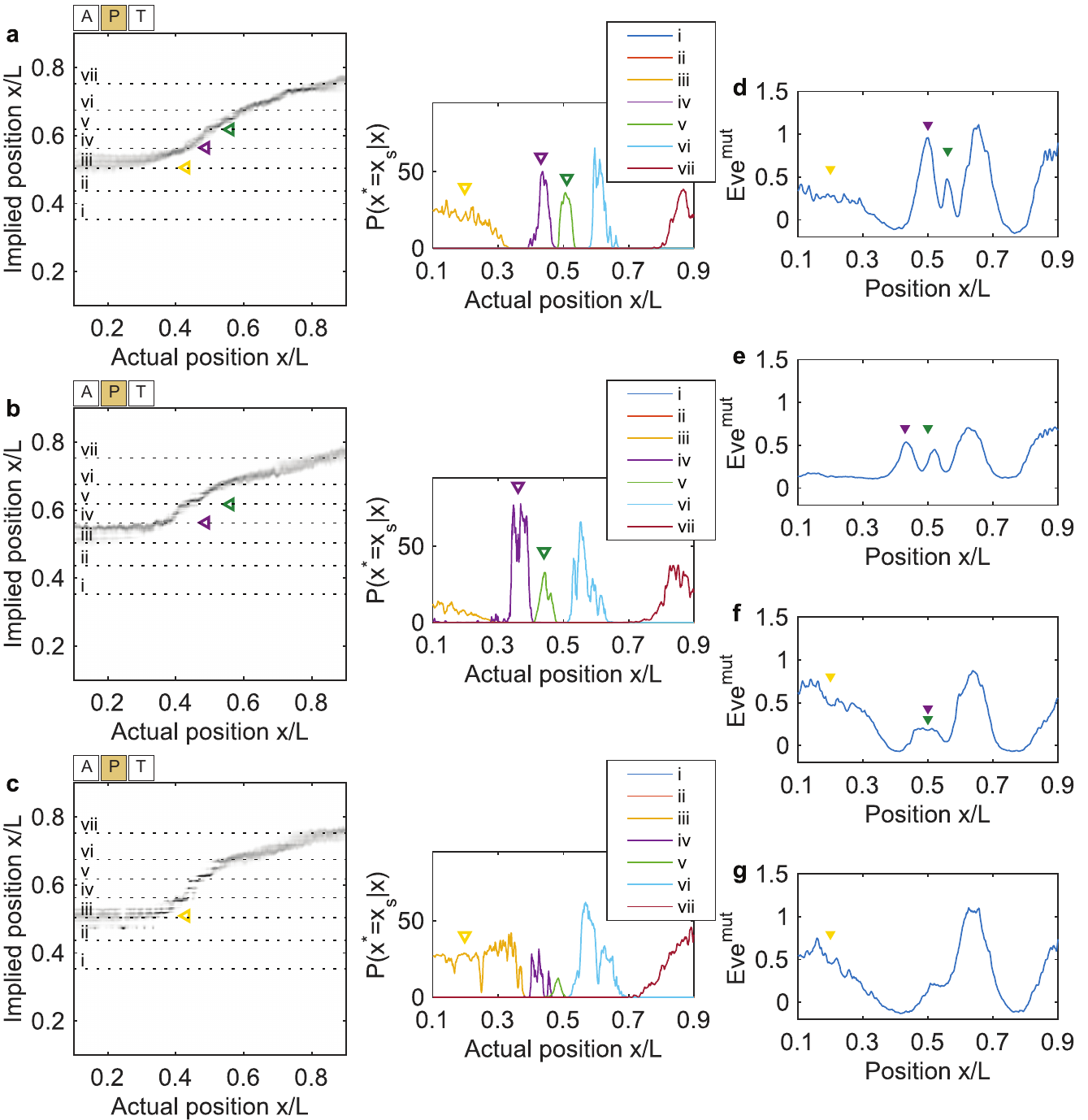}}
\hspace{1cm}
\caption{{\bf Predicting variable number of Eve stripes in $bcd$ $tsl$ mutants.} {\bf a.-c.} Decoding map from individual $bcd$ $tsl$ embryo. Horizontal dashed lines indicate the average locations of Eve peaks, their intersections with the decoding map are shown in the side panel (right, $P(x^*|x)$). Stripes {\it iv} and {\it v} (purple and green open triangles, respectively), and diffuse stripe {\it iii} (yellow open triangle) are predicted to have variable expressivity: {\bf a.} all stripes are predicted, {\bf b.} diffuse stripe {\it iii} is missing, {\bf c.}  stripes {\it iv,v} are either overlapping or missing. {\bf d-g.} We find examples of such variability in the measured Eve expression profiles from $bcd$ $tsl$ mutants: {\bf d}, embryo with both {\it iv} and {\it v} stripes (purple and green filled triangles, respectively), and diffuse stripe {\it iii} (yellow filled triangle), {\bf e}, embryo with diffuse stripe {\it iii} missing, {\bf f}, embryo with overlapping stripes {\it iv,v}, and {\bf g}, embryo with either {\it iv} and {\it v} missing. \label{SF14}}
\end{SCfigure*}

\begin{SCfigure*}
\hspace{1.5cm}
\centerline{\includegraphics[width =110mm]{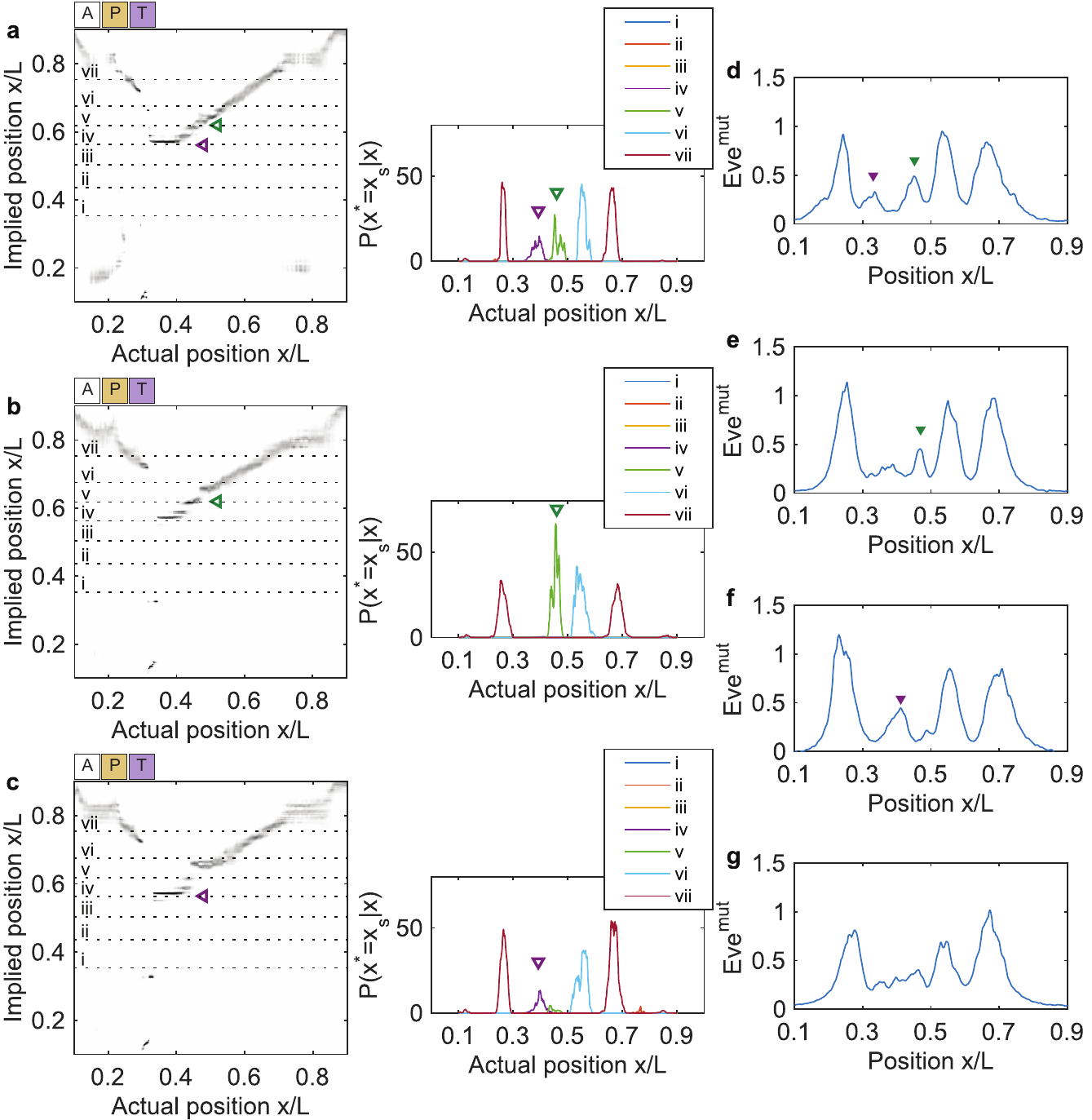}}
\hspace{1cm}
\caption{{\bf Predicting variable number of Eve stripes in $bcd^{E1}$ mutants.} {\bf a.-c.} Decoding map from individual $bcd^{E1}$ embryo. Horizontal dashed lines indicate the average locations of Eve peaks, their intersections with the decoding map are shown on the side panel (right, $P(x^*|x)$). Stripes {\it iv} and {\it v} (marked by purple and green open triangles, respectively) are predicted to have variable expressivity: {\bf a.} both stripes are predicted, {\bf b.} only stripe {\it v} is predicted, {\bf c.} only stripe {\it iv} is predicted. {\bf d-g.} We measure such variability in the number of expressed Eve stripes in $bcd^{E1}$ mutants: {\bf d}, embryo with both {\it iv} and {\it v} stripes (purple and green filled triangles, respectively), {\bf e}, embryo with {\it iv} missing, {\bf f}, embryo with {\it v}, missing, and {\bf g}, embryo with both {\it iv} and {\it v} missing. \label{SF13}}
\end{SCfigure*}

\begin{SCfigure*}
\hspace{1.5cm}
\centerline{\includegraphics[width = 120mm]{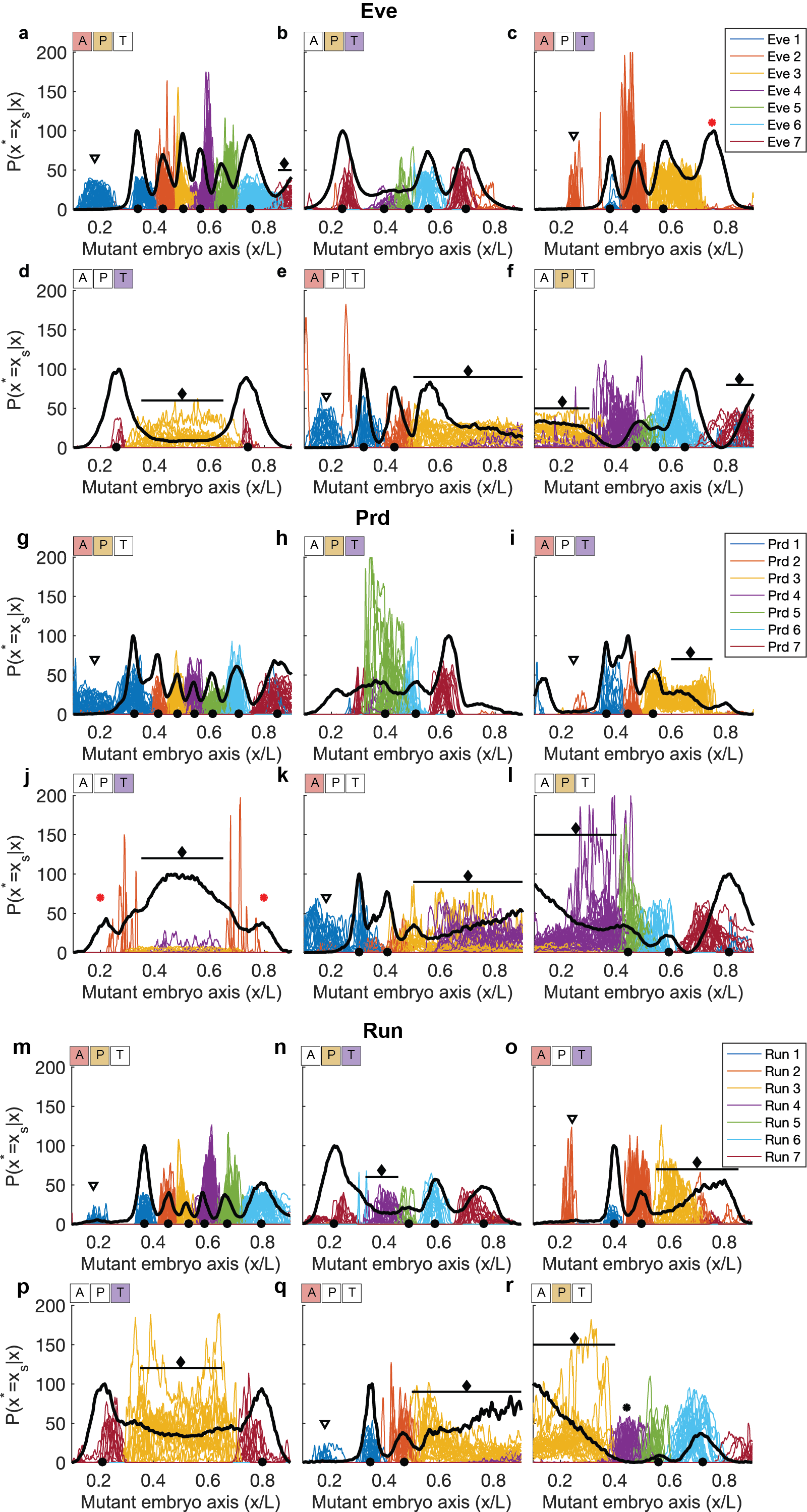}}
\hspace{1.75cm}
\caption{{\bf Annotated stripe predictions in mutant embryos. } Horizontal intersections of probability density $P(x^*|x)$  maps in individual mutant embryos at the average locations of wild-type pair rule stripes.  {\bf a-f.} Eve peaks, {\bf j-l.} Prd peaks, {\bf m-r.} Run peaks. For reference, we show the mean pair-rule expression for each mutant background (solid black line), arbitrary units scaled to match the y-axis limits. Filled black circles on the x-axis mark the average locations of measured peaks, which are successfully predicted from the decoding maps  and plotted in Fig.~\ref{summary}. Predicted diffuse stripes are marked by filled diamonds over horizontal lines, which span the diffuse stripes. Open triangles show ``echoes" as in Fig.~\ref{SF15}.  Interestingly, a duplication of Eve stripe 7, and diffuse expression of stripes 3-4 are found expressed where predicted in the anterior ({\bf f}). Red stars shows observed, but not predicted Eve stripe ({\bf c}), and Prd stripes ({\bf j}). Black star shows predicted, but not observed Eve stripe ({\bf f}), and Run stripe ({\bf r}). \label{SF16Eve}}
\end{SCfigure*}

\begin{figure*}[t]
\centerline{\includegraphics[width = 0.75\linewidth]{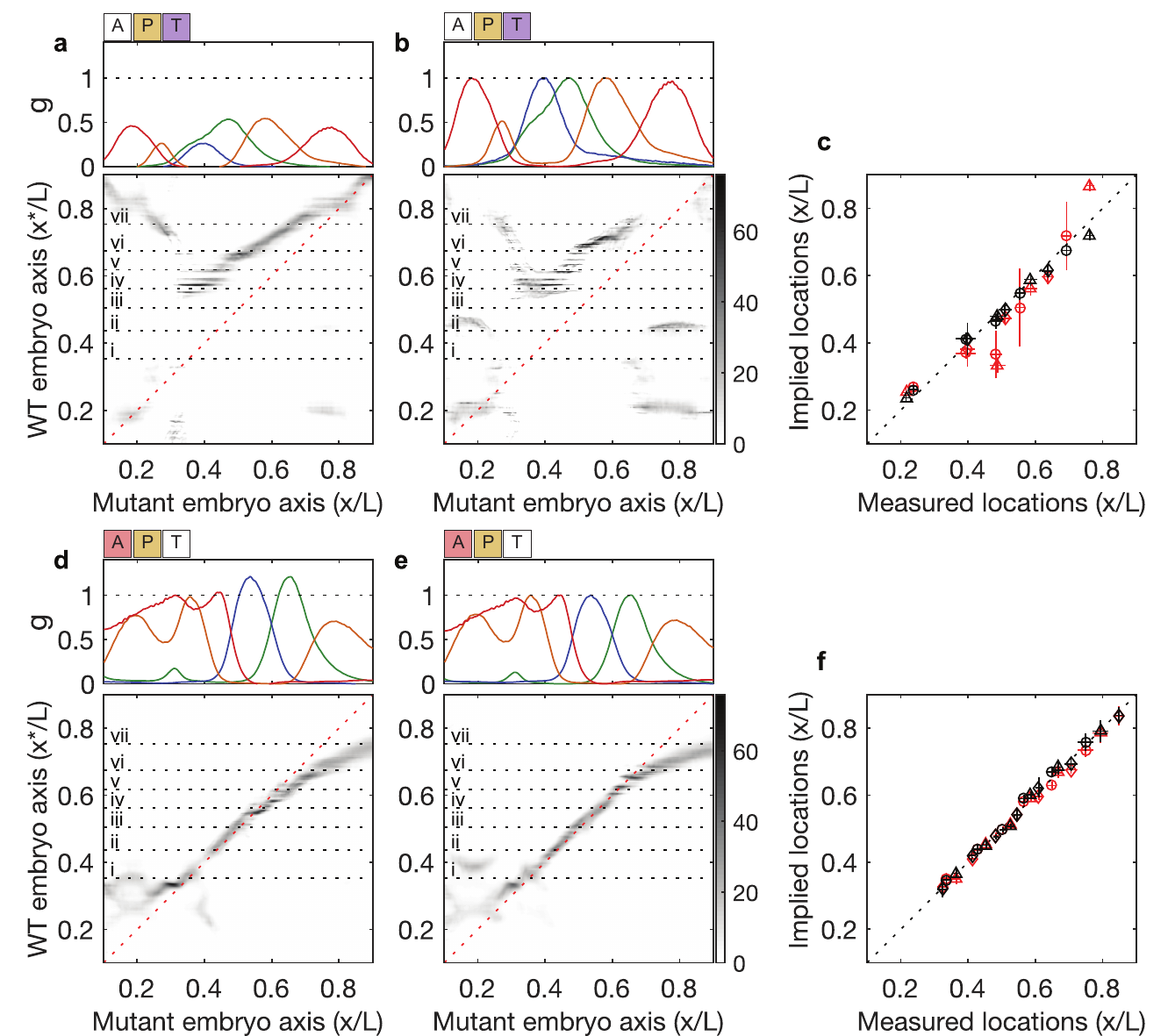}}
\caption{{\bf Absolute expression levels matter for decoding.} We predict pair-rule stripes in $bcd^{E1}$ embryos, whose gap gene expression is absolute units, normalized to reference wild-type embryos ({\bf a.}), or normalized with respect to themselves ({\bf b.}); {\bf d.} and  {\bf e.}, as in {\bf a.} and  {\bf b.} but  for $etsl$ embryos. Top panels, mean gap gene expression in respective units; bottom panels, average decoding map with horizontal dotted lines at the average locations of wild-type Eve stripes (roman numerals).   {\bf c.}   and {\bf f.} Summary of stripe predictions from decoding based on absolute (red) or normalized (black) expression levels.  
 \label{SF10}}
\end{figure*}

Decoding in individual embryos permits us to compare not only the predicted mean stripe locations to the measurements, but also to compare the variability in stripe position, shape, and in the total number of observed stripes. Figures~\ref{SF14} and \ref{SF13} show examples of individual Eve profiles where some of the Eve stripes 3, 4, 5 were either missing or had a broad, poorly localized ``diffuse'' profile in mutant backgrounds. We predicted these phenomena for exactly the same mutant backgrounds and the same stripes from individual embryo decoding maps.

A detailed description of individual embryo pair-rule stripe predictions in mutant backgrounds, analogous to Fig~\ref{SF15} for the wild-type, is shown in Fig~\ref{SF16Eve}. In these panels, we denote separately diffuse stripes, as well as a small number of observed-but-not-predicted and predicted-but-unobserved stripes. All non-diffuse predictions across the three pair-rule genes and all mutants are shown in the summary Fig~\ref{F4} in the main text.

We invested substantial experimental effort to measure gap gene expression levels in mutant embryos side-by-side with the wild-type controls, so that  absolute concentrations   can contribute to the decoding. But do they? In Fig~\ref{SF10} we show the effect of the absolute level on the decoding map, and consequently on the pair-rule stripe prediction performance. In the {\em bcd} mutant background (Fig~\ref{SF10}a,b), gap gene expression levels are strongly perturbed in shape but also suppressed in magnitude by  $\sim 2\times$. Decoding these profiles gives predictions of pair-rule stripes that agree very closely with data (Fig~\ref{SF10}c, black symbols). In contrast, when mutant profiles are individually normalized so that they span the range of expressions between 0 and 1---in essence, keeping the profile shape but undoing the magnitude effect---leads to much worse predictions of pair-rule stripes (Fig~\ref{SF10}c, red).

In the {\em tsl} mutant background, the effect of absolute concentrations is more subtle.  In these mutants, Kr and Kni are overexpressed by $\sim 10-20\%$ relative to the wild-type, which leads to a slight deformation in the decoding map in the posterior ($x>0.5$), and this effect disappears if we normalize to keep only relative expression levels. While the effect is smaller than in the {\em bcd} background,  pair-rule stripes at $0.6 < x < 0.7$ are consistently predicted better using absolute gap gene concentrations. In sum, both for large scale and precision effects on our pair-rule predictions, being able to measure gap gene concentrations relative to the wild-type is crucial. This suggests as well that the embryo itself responds to precisely determined, absolute concentrations of signaling molecules.

\begin{figure}
\centerline{\includegraphics[width =  0.8\linewidth]{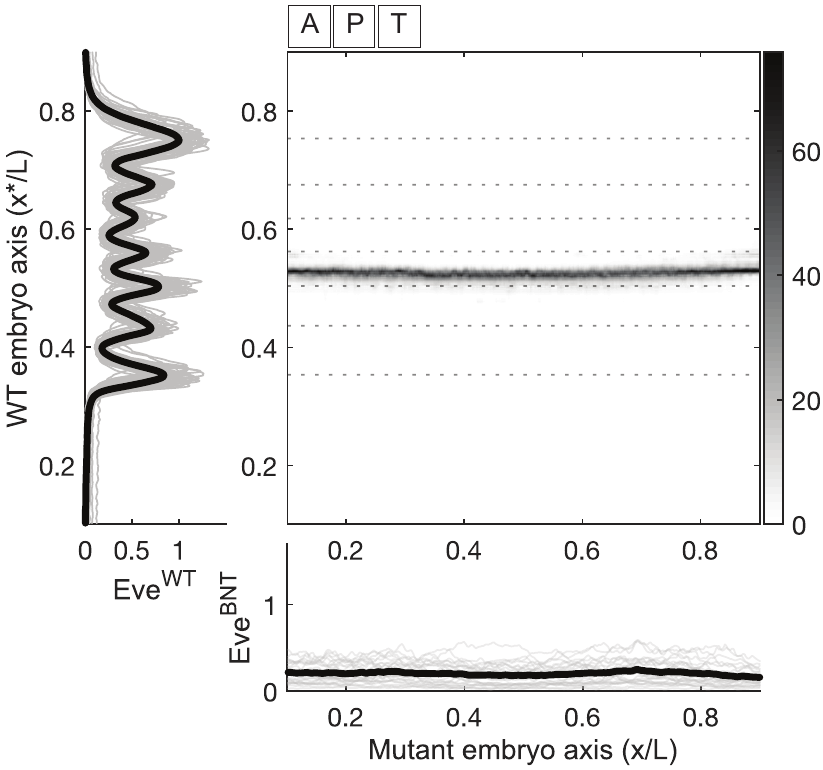}}
\caption{{\bf Deleting three maternal inputs abolishes positional information. } Decoding map for the triple deletion mutant {\em bcd, osk, tsl}.   Positions of Eve stripes in the wild-type (left) fail to intersect the map, consistent with the absence of stripes in the mutant (bottom). \label{flatline}}
\end{figure}

When we delete all three of the maternal inputs, positional information is abolished (Fig \ref{flatline}).  There is a  low uniform expression of Eve in mutant embryos,  consistent with the predictions from our decoding map. 

\end{document}